# Demonstration of low emittance in the Cornell energy recovery linac injector prototype

Colwyn Gulliford,* Adam Bartnik, Ivan Bazarov,† Luca Cultrera, John Dobbins, Bruce Dunham,
Francisco Gonzalez, Siddharth Karkare, Hyeri Lee, Heng Li, Yulin Li, Xianghong Liu, Jared Maxson,
Christian Nguyen, Karl Smolenski, and Zhi Zhao

*Cornell Laboratory for Accelerator-Based Sciences and Education (CLASSE), Cornell University, Ithaca, New York 14853, USA*
(Received 9 April 2013; published 16 July 2013)

We present a detailed study of the six-dimensional phase space of the electron beam produced by the Cornell Energy Recovery Linac Photoinjector, a high-brightness, high repetition rate (1.3 GHz) DC photoemission source designed to drive a hard x-ray energy recovery linac (ERL). A complete simulation model of the injector has been constructed, verified by measurement, and optimized. Both the horizontal and vertical 2D transverse phase spaces, as well as the time-resolved (sliced) horizontal phase space, were simulated and directly measured at the end of the injector for 19 and 77 pC bunches at roughly 8 MeV. These bunch charges were chosen because they correspond to 25 and 100 mA average current if operating at the full 1.3 GHz repetition rate. The resulting 90% normalized transverse emittances for 19 (77) pC/bunch were $0.23 \pm 0.02$ ($0.51 \pm 0.04$) $\mu$m in the horizontal plane, and $0.14 \pm 0.01$ ($0.29 \pm 0.02$) $\mu$m in the vertical plane, respectively. These emittances were measured with a corresponding bunch length of $2.1 \pm 0.1$ ($3.0 \pm 0.2$) ps, respectively. In each case the rms momentum spread was determined to be on the order of $10^{-3}$. Excellent overall agreement between measurement and simulation has been demonstrated. Using the emittances and bunch length measured at 19 pC/bunch, we estimate the electron beam quality in a 1.3 GHz, 5 GeV hard x-ray ERL to be at least a factor of 20 times better than that of existing storage rings when the rms energy spread of each device is considered. These results represent a milestone for the field of high-brightness, high-current photoinjectors.

## I. INTRODUCTION

The desire for light sources with substantially more coherence and brightness has fueled significant interest in the research and design of energy recovery linacs (ERLs) and free electron lasers (FELs). The feasibility of ERL technology has already been demonstrated at several laboratories, most notably Thomas Jefferson National Accelerator Facility (TJNAF) [1], where energy recovery was achieved for 100 MeV beams with an average current of up to 9 mA. However, in order to design and construct a large scale, high energy (GeV) ERL x-ray source, significant advancement of both superconducting rf (SRF) cavity technology, as well as high-brightness, high-current sources has been required. Over the past several years, Cornell University has played a lead role in the development of both areas, and has successfully reached several major milestones towards the realization of a practical ERL x-ray facility [2].

To drive this type of machine requires an exceptional electron source producing high-brightness bunches at high repetition rates. Traditionally it has been thought that the best beam quality was obtained using low duty factor normal conducting rf (NCRF) gun based photoemission sources [3–5], as these devices are capable of providing high peak cathode fields. These fields are typically in the vicinity of 100 MV/m, though the field at the cathode during emission is often significantly lower since these devices are usually run off-crest [4,5]. Because of the considerable heat load generated in the cavity walls, the cw operation of NCRF sources requires substantial lowering of the electric gradient, an approach being pursued at several facilities [3,6]. Work started at TJNAF, and later expanded at Cornell University, shows that the combination of a high-voltage DC gun followed immediately by acceleration with superconducting cavities yields beams with single bunch quality rivaling that produced by rf guns, but at much higher (GHz) repetition rates [7,8]. In addition, DC guns provide an excellent vacuum, allowing for a much wider range of cathode materials to be used than in NCRF guns. While SRF guns show significant promise for producing high-brightness, high-current beams, this technology is currently in the development and testing stage, and the achieved beam parameters so far are relatively modest [9].

*cg248@cornell.edu
†ib38@cornell.edu

Consequently, a photoinjector using a DC gun has been designed, built, and commissioned at Cornell University. One of the main goals of this project was to produce high average current from this source. The Cornell injector has made great strides toward this end, having recently set several new records for high average current from a photoinjector with cathode lifetime suitable for an operating facility [10,11]. Another major goal is the demonstration of low emittance at the end of the injector's merger section, where the (relatively) low energy beam would be injected into the main ERL linac. The results in this work demonstrate that it is possible to produce and transport beams from a DC source which have emittances at the point of injection approaching the diffraction limit for hard x rays, and which have a bunch length and an energy spread within the parameter space required by the specifications of a full hard x-ray ERL.

In general, to achieve the maximum brightness in a photoinjector, it is crucial to control both the transverse and longitudinal space charge forces, as well as the effects of time-dependent rf focusing [8,12–16]. Effective emittance compensation is possible when bunches are created with a charge distribution that has predominantly linear space charge fields [12,14–16], and if done correctly, can lead to final emittances approaching the intrinsic emittance of the photocathode. One fundamental limit to this approach occurs when the amount of charge extracted from the cathode nears the virtual cathode instability limit. A rough calculation shows that the lowest achievable emittance then becomes proportional to the square root of the bunch charge $q$ [16]:

$$\epsilon_n \propto \sqrt{q \cdot \frac{\text{MTE}}{E_{\text{cath}}}}. \qquad (1)$$

Here MTE and $E_{\text{cath}}$ are the mean transverse energy of the photoelectrons and the accelerating field at the cathode, respectively. Detailed simulations of well optimized DC gun photoinjectors support this square root dependence on the bunch charge and the cathode's MTE [7,8]. In this paper, we show that the final measured emittance also scales in accordance with Eq. (1). This represents a key step in experimentally realizing the maximum brightness limit for photoinjectors.

The outline of this work is structured as follows. First, a general description of the Cornell ERL photoinjector is given. This includes a description of the beam line layout, the relevant accelerating and optical elements, and the diagnostic systems used to take our emittance data. Next, we describe how to model the dynamics in the injector using the space charge simulation code GENERAL PARTICLE TRACER (GPT) [17], and give a verification of the GPT injector model against linear optics measurements. After this, a description of the optimization of this model and the process for determining our final optics settings used in the experiment is given. This is followed by the main results of this work. These include direct measurement and simulation of both the projected transverse phase spaces, as well as the time-resolved horizontal phase space at the end of the injector merger section. Additionally, the energy spread distribution was measured using a single dipole magnet in a separate diagnostic beam line section, providing an upper bound on the rms energy spread at the end of the merger.

## II. THE CORNELL ERL INJECTOR

Construction of the Cornell injector was completed in the summer of 2007. Initial beam commissioning experiments revealed an issue with charging up of the ferrites in the higher-order mode dampers in the injector cryomodule. After this problem was successfully addressed [18], beam experiments started in earnest in the spring of 2010 and have continued to this date [10,11,19–25]. In that time, significant progress towards meeting the target goals of the injector project has been made. Table I shows these specifications. Of particular interest to this work are the specifications for the normalized transverse emittance and rms bunch length. We demonstrate later in this work that these specifications have been met.

### A. Description and layout

The layout of the Cornell ERL injector is shown in Fig. 1. The Cornell injector features two laser systems. The primary system is a 1.3 GHz laser producing 520 nm, 1 ps rms pulses with an average power of up to 60 W [26], and is used for high-current experiments. For emittance measurements with nonzero bunch charge, we exclusively use a 50 MHz system, whose individual pulses have comparable pulse energy and duration to the 1.3 GHz laser. This laser system allows us to limit the average electron beam power hitting our interceptive emittance diagnostics. After being generated in one of these two lasers, the final laser pulse train can be chopped using a Pockels cell, and shaped using our temporal shaping system [20]. This system consists of four rotatable birefringent crystals, which are used to divide the primary laser pulse into 16 copies, with tunable relative intensities set by their rotation angles.

TABLE I. List of injector design specifications and target parameters.

| Parameter | Specification |
|---|---|
| Beam energy | 5–15 MeV |
| Normalized emittance | $\epsilon_n \leq 0.3\ \mu$m |
| rms bunch length | $\sigma_t \leq 3$ ps |
| Bunch charge | 77 (19) pC |
| Average current | 100 (25) mA |

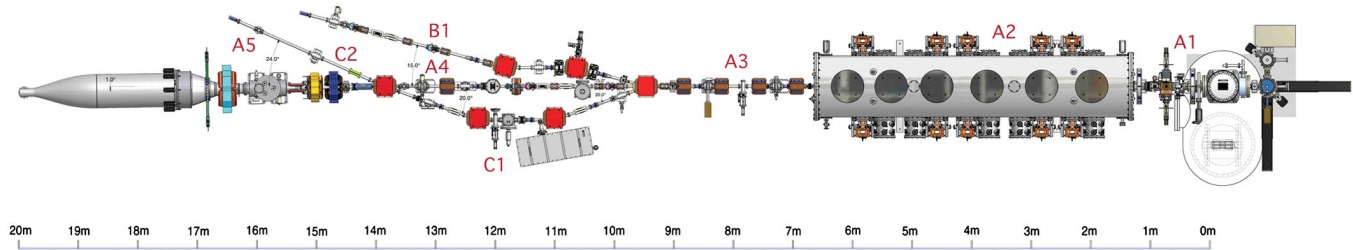

FIG. 1. Top view of the Cornell ERL injector.

These crystals are typically tuned to produce a roughly flat intensity profile, with around 8 ps rms duration.

The cathode used for this study was a GaAs wafer grown using molecular beam epitaxy on a $p$-doped GaAs substrate. The cathode was heat cleaned to 620°C for 2 hours and then activated to negative electron affinity using Cs and $NF_3$ via the "yo-yo" process. The doping density was $5 \times 10^{18}$ cm$^{-3}$. The top 100 nm was left undoped. The resulting cathode had a quantum efficiency of 4%, a mean transverse energy of 90 meV, and a subpicosecond response time at 520 nm.

The high-voltage DC gun used in these measurements is the same one used in previous space charge and emittance studies [10,11,19–25]. The gun was operated at 350 kV for all measurements in this work. The beam line section just after the gun, labeled "A1" in Fig. 1, houses two emittance compensation solenoids and a 1.3 GHz normal conducting buncher cavity. These elements were used to compensate the initial emittance blowup near the cathode, and to compress the bunch longitudinally before further acceleration. Immediately after emittance compensation, the bunches were accelerated using the five superconducting niobium cavities in the SRF cryomodule, labeled "A2" in Fig. 1. In addition to increasing the beam energy, and thus partially freezing in the emittance, the SRF cavities were also used to perform further emittance compensation and longitudinal compression via time-dependent transverse and longitudinal focusing. Each cavity features a symmetric twin input coupler design in order to eliminate any time-dependent dipole kick [27,28] and can be operated with a voltage in the range of 1 to 3 MV. For a more detailed description of the injector cavities see [29].

Just after the cryomodule, the beam was passed through a four-quad telescope, labeled "A3" in Fig. 1. The beam was then directed into one of several diagnostic beam line sections. The section most relevant to this work is the "B1" merger shown in detail in Fig. 2. The injector merger section is comprised of a conventional three-dipole achromat [30–32]. This design was chosen for its simplicity, and due to the limited space available for the injector experiment. The trade off for this approach is that while this merger setup closes the single particle dispersion, it does not satisfy the second achromat condition $\eta'_{sc} = 0$ for the space charge dispersion function [30,31]. Despite this, both our simulations and measurements show that this merger design does in fact preserve low emittance for our operating parameters. As was anticipated in [31], this was accomplished by finding the correct settings for the four quadrupole magnets in the A3 straight section.

The emittance measurement system (EMS) used for projected and time-resolved phase-space measurements is a two-slit system with no moving parts [22]. Figure 2 shows the layout of this diagnostic system. In front of each 20 $\mu$m slit is a scanner magnet. Each scanner magnet consists of a pair of air core correcting coils with equal and opposite field polarity and negligible sextupole field component. The resulting effect of the scanner magnet is to translate the beam transversely without imparting any angle to it. In practice, the coil pairs in each scanner magnet cancel each other to better than a few percent [22]. For projected phase-space measurements, the beamlet coming through both slits was collected using the Faraday cup at the end of the merger section. For time-resolved horizontal phase-space measurements, the beamlet was passed through a horizontal deflecting cavity [33] in order to resolve the time axis of the beam on the viewscreen at the end of the merger section [25]. For a more detailed description of the EMS, refer to [22,25].

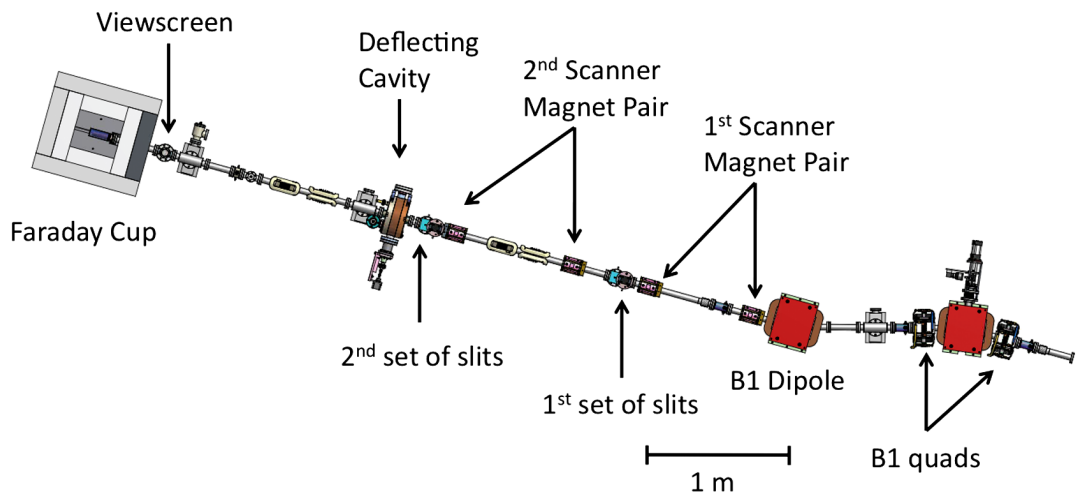

FIG. 2. Top view of the B1 injector merger section showing the emittance measurement system.

### B. The GPT injector model

The 3D space charge code GPT was used extensively in this work. To model space charge effects, GPT utilizes a 3D nonequidistant mesh solver [34,35]. Additionally, GPT allows users to define their own custom optical elements, as well as position and superimpose electromagnetic field maps in 3D space. These features provided sufficient versatility to accurately model our machine, where the fields of several elements overlap. All of the beam line elements relevant for the space charge simulations in this work have been modeled using realistic field maps.

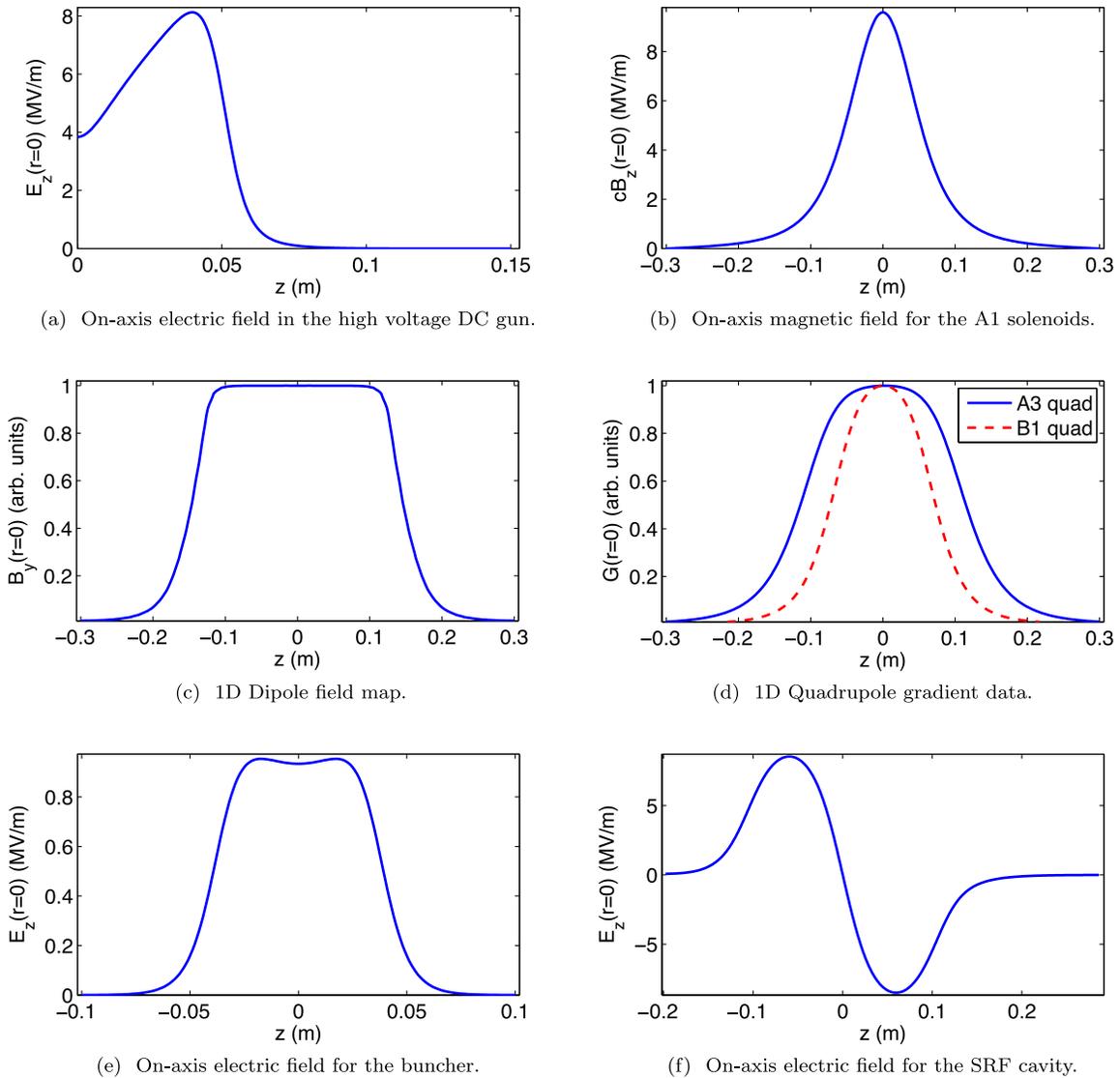

FIG. 3. On-axis electric and magnetic fields for (a) the high-voltage DC gun at 350 kV, (b) emittance compensation solenoid at 3.75 A, (c) the A3 and B1 merger dipoles, (d) the A3 and B1 merger quads, (e) the buncher cavity at 60 kV, and (f) the SRF cavity at 1 MV.

POISSON-SUPERFISH [36] was used to generate 2D cylindrically symmetric fields specifying $E_r(r, z)$ and $E_z(r, z)$, as well as $B_r(r, z)$ and $B_z(r, z)$, for the high-voltage DC gun and emittance compensation solenoids, respectively. The on-axis fields for these elements are shown in Figs. 3(a) and 3(b).

In order to efficiently and accurately describe the injector dipoles and quadrupoles, we created custom GPT elements which generate 3D fields using an off-axis field expansion of 1D field data. To create the 1D dipole and quadrupole field data, the full 3D fields for each type of element were computed in OPERA-3D [37]. From these fields the quantities $B_y(r=0, z)$ and $\partial B_y(r=0, z)/\partial x$ were extracted from the dipole and quadrupole fields, respectively. Figures 3(c) and 3(d) show the 1D field data used for the dipoles and quadrupoles in the injector. Our custom GPT rectangular dipole element uses an off-axis expansion of the fields given by [38]

$$B_x \sim \mathcal{O}(4), \qquad B_y = B_0(z) - \frac{y^2}{2}\frac{d^2 B_0}{dz^2} + \mathcal{O}(4),$$
$$B_z = y\frac{dB_0}{dz} + \mathcal{O}(4), \qquad (2)$$

to model the higher-order components of the dipole field. In this expression $B_0 = B_y(r=0, z)$. This expansion assumes that the particles do not see the fringe fields on the lateral sides of the magnet. This is true for the particle trajectories and magnets in the injector, where the maximum simulated rms beam size through the dipoles was $\leq 3$ mm [see Fig. 13(b)], and the dipole width was 25 cm. Similarly, the fields for the quadrupoles were computed with an off-axis field expansion [38]:

$$B_x = y\left[G(z) - \frac{1}{2}(3x^2 + y^2)\frac{dG}{dz}\right] + \mathcal{O}(5),$$
$$B_y = x\left[G(z) - \frac{1}{2}(3y^2 + x^2)\frac{dG}{dz}\right] + \mathcal{O}(5), \qquad (3)$$
$$B_z = xyG(z) + \mathcal{O}(4).$$

Here the term $G(z) = \partial B_y/\partial x(r=0, z)$. To verify Eqs. (2) and (3), single particle tracking through the fields created by our custom elements was compared to tracking using the full 3D field maps. Excellent agreement was found in both cases. Additionally, the custom elements proved significantly faster because they do not require look-up of 3D field arrays.

All rf cavity fields were generated using the eigenmode 3D field solver in CST MICROWAVE STUDIO (MWS) [39]. The buncher cavity was modeled using a 2D cylindrically symmetric map specifying $E_r(r, z)$, $E_z(r, z)$, and $H_\theta(r, z)$. The on-axis field map for the buncher is shown in Fig. 3(e).

Previous work demonstrates that asymmetric focusing of the bunch near the input power couplers of the accelerating cavities is significant and can lead to asymmetric horizontal and vertical emittances [25]. To address this issue, we generated full 3D field maps for the accelerating cavities which incorporate the beam running conditions following our method outlined in [40]. Figure 4 shows the 3D cavity model used in MWS for the accelerating cavities. The procedure for correctly constructing the fields in the coupler and cavity requires two sets of MWS solutions. Each set of fields was created by terminating the input coupler line in the MWS model with either an electric or magnetic wall boundary condition [40]. From these solutions, traveling waves carrying power into and out of the cavity through the couplers were constructed, scaled, and shifted in phase to match the actual running conditions in the injector. In order to further limit the beam power deposited in our interceptive EMS, the pulse train from the 50 MHz laser was chopped using a Pockels cell. The resulting beams typically had currents on the order of a microamp or less. In generating the field maps for the accelerating cavities, this amounts to effectively having zero current. The parameter which determines how the cavity fields depend on the operating conditions is the reflection coefficient [40]:

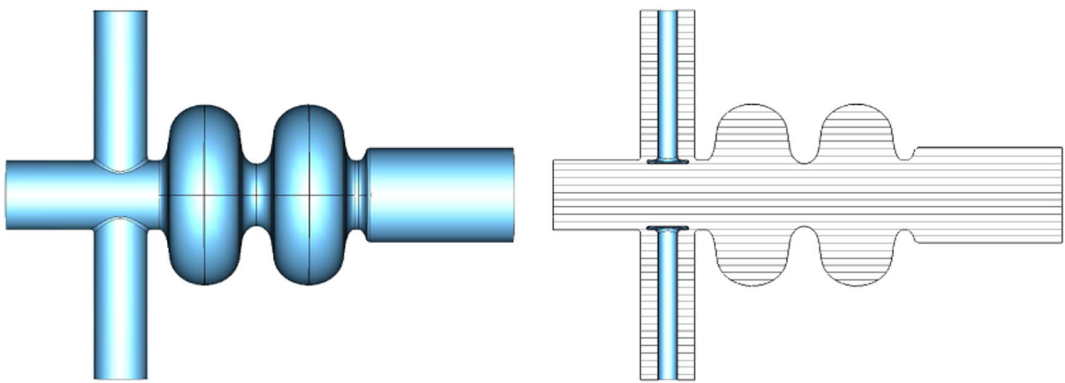

(a) Exterior view of the MWS injector cavity model. (b) Cutaway view of the MWS injector cavity model.

FIG. 4. (a) The MICROWAVE STUDIO model of the ERL injector cavity: (a) the cavity and coupler exterior, (b) cutaway view of the same model showing the inner coupler antennas.

$$\Gamma = -\frac{\frac{1-\beta}{1+\beta} + \frac{I_b}{V_c}(R/Q)Q_L e^{-i\phi_0} + i\tan\psi'}{1 + \frac{I_b}{V_c}(R/Q)Q_L e^{-i\phi_0} + i\tan\psi'}. \quad (4)$$

Here $I_b$ and $\phi_0$ are the average beam current and phase of the beam with respect to the cavity fields. The rest of the parameters in this expression describe the properties of the cavities: $\beta$ is the coupling parameter, $Q$ is the intrinsic cavity quality factor, $R$ is the shunt resistance, $V_c$ is the cavity voltage, $Q_L = Q_0/(1+\beta)$ is the loaded quality factor, and $\tan\psi'$ is the cavity detuning parameter. In the zero current limit (with the cavity tuned to resonance), the reflection coefficient reduces to

$$\Gamma(I_b \to 0) = \frac{\beta - 1}{\beta + 1}. \quad (5)$$

This implies that the fields in the cavity and coupler coax depend only on the amount of coupling. The coupling factor $\beta$ is determined by how far the inner coupler antennas are retracted from being flush with the beam pipe. For our emittance measurements, the couplers were fully retracted (zero current setting). By fully retracting the coupler antennas in the MWS model, and generating two sets of solutions for both boundary conditions on the end of the coupler coax, we created one set of complex 3D electric and magnetic field maps for the SRF cavities. Figure 3(f) shows the resulting on-axis electric field. We point out that simulations subsequently showed that the asymmetric emittances caused by the rf quad effect in the cavities could be successfully remedied by appropriate choice of magnetic quadrupole focusing downstream.

After completing the GPT physics model of the injector, we developed a user interface between the real machine and its GPT counterpart. Named the "virtual accelerator GUI," this program was designed to provide a single interface between the corresponding optics settings in the EPICS control system of the real machine, and the stand alone GPT code. Additional features include the ability to save and load optics settings and simulation results to and from file, the ability to load injector settings from the machine and independently adjust them in simulation, as well as the ability to visualize all relevant simulation data. A screen shot of this application is shown in Fig. 5. In constructing this program, a master GPT input file was created which included not only the optical elements described in this section, but also simulation output screens at all of the corresponding locations of the beam position monitors (BPMs), viewscreens, and emittance measurement systems in the injector. The result was a nearly one-to-one simulation counterpart to the real machine. With this, we were able to use GPT in a more useful and realistic way, with simulations often guiding experiments in near real-time in the control room.

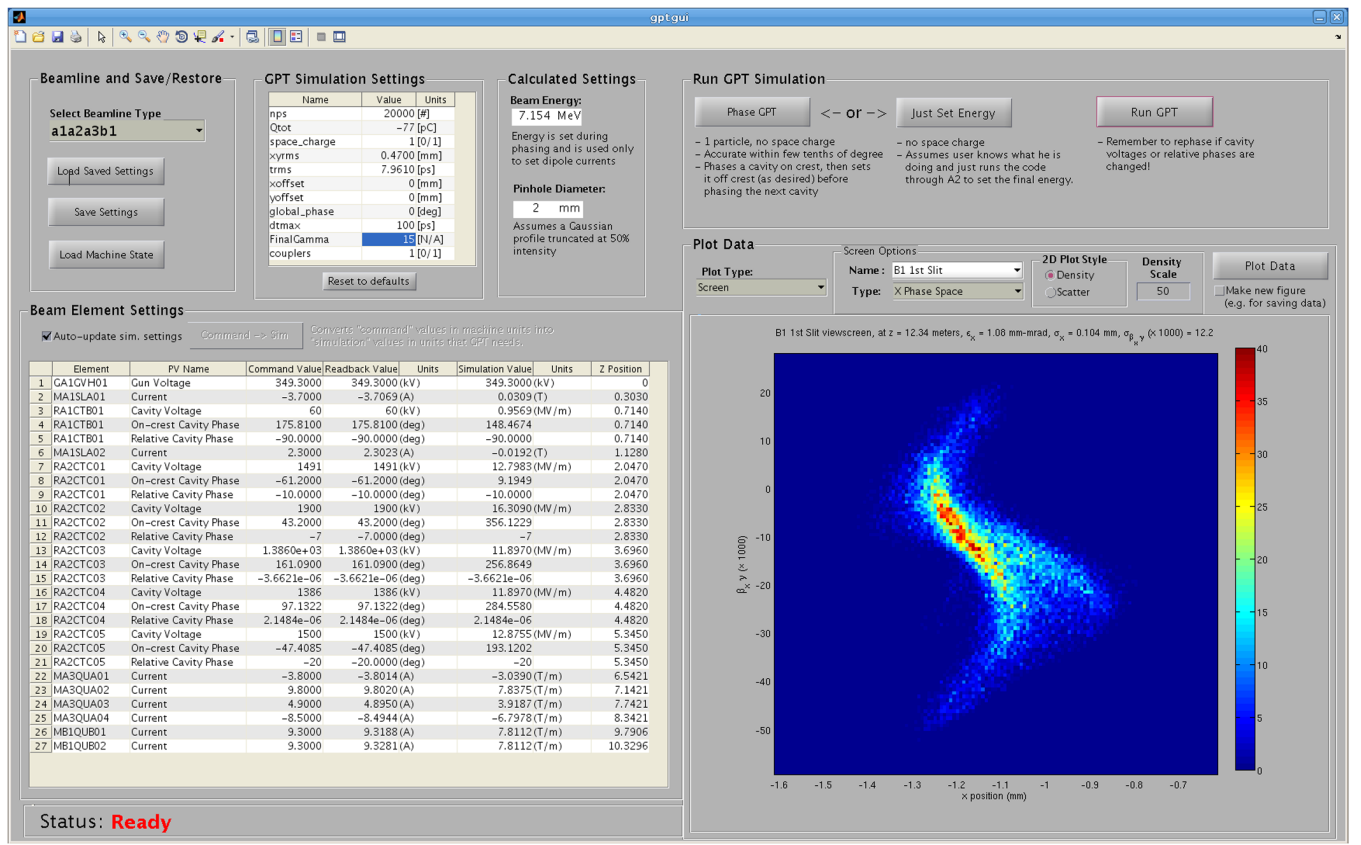

FIG. 5. Screenshot of the virtual accelerator GUI.

## III. MEASUREMENTS

All of the measurements in this work fall into one of two categories: measurements performed at near-zero bunch charge for verification and calibration purposes; or phase-space measurements of space charge dominated bunches.

### A. Measurements at near-zero bunch charge

The measurements presented in this section include comparison of difference orbits (linear optics) in the injector with the GPT model including the effects of the rf input couplers, measurement of the beam size envelope along the injector and its verification with simulation, and calibration of the EMS and analysis procedures by comparing the emittance computed from the direct measurement of the projected transverse phase spaces in the merger and the emittance measured using a solenoid scan in the gun vicinity. In order to accurately perform difference orbit measurements, the BPM system needed to be corrected for its nonlinear response, the procedure for which is presented below.

#### 1. BPM correction procedure

The injector BPMs consist of four striplines, as seen in Fig. 6(a). To model the system we make two assumptions: (i) both the beam pipe and striplines are assumed to be infinitely long perfect conductors connected to ground; and (ii) the beam is taken to be an infinite line charge at the position $\mathbf{r}_b = (x_b, y_b)$. The first assumption implies that the potential must vanish at the beam pipe. This is accomplished by placing an image line charge with opposite charge density at $\mathbf{r} = (R^2/r_b^2)\mathbf{r}_b$, where $R$ is the beam pipe radius. The resulting electric field everywhere is

$$\mathbf{E}(\mathbf{r}, \mathbf{r}_b) = \frac{\lambda}{2\pi\epsilon_0}\left[\frac{\mathbf{r} - \mathbf{r}_b}{|\mathbf{r} - \mathbf{r}_b|^2} - \frac{\mathbf{r} - (R^2/r_b^2)\mathbf{r}_b}{|\mathbf{r} - (R^2/r_b^2)\mathbf{r}_b|^2}\right].$$

From the electric field, the surface charge density on the beam pipe and striplines can be computed using $\sigma = -\epsilon_0 \mathbf{E}(\mathbf{R}, \mathbf{r}_b) \cdot \hat{\mathbf{n}}$, where $\hat{\mathbf{n}}$ is the normal vector to the beam pipe surface. Since this model assumes the beam pipe is a perfect conductor, the field is perpendicular to the surface so that $\mathbf{E}(\mathbf{R}, \mathbf{r}_b) \cdot \hat{\mathbf{n}} = |\mathbf{E}(\mathbf{R}, \mathbf{r}_b)|$. In cylindrical polar coordinates, the surface charge density takes the form

$$\sigma(R, \theta, r_b, \theta_b) = \frac{\lambda}{2\pi}\left(\frac{(r_b^2/R^2) - 1}{R^2 + r_b^2 - 2Rr_b\cos(\theta - \theta_b)}\right)R.$$

The angles $\theta$ and $\theta_b$ are defined in Fig. 6(a). The signal from the $i$th stripline is defined as the fraction of the surface charge density found on that stripline:

$$S_i(x_b, y_b) = \frac{1}{\lambda}\int_{\theta_i - \theta_s/2}^{\theta_i + \theta_s/2} \sigma(R, \theta, x_b, y_b) R d\theta. \quad (6)$$

Here the angle $\theta_s$ is the angle subtended by each stripline. Performing the integration yields

$$S_i(x_b, y_b) = -\frac{1}{\pi}\tan^{-1}\left[\left(\frac{R + r_b}{R - r_b}\right)\tan\left(\frac{\theta - \theta_b}{2}\right)\right]\bigg|_{\theta_i - \theta_s/2}^{\theta_i + \theta_s/2}, \quad (7)$$

where $\theta_i \in \{0, \pi/2, \pi, 3\pi/2\}$.

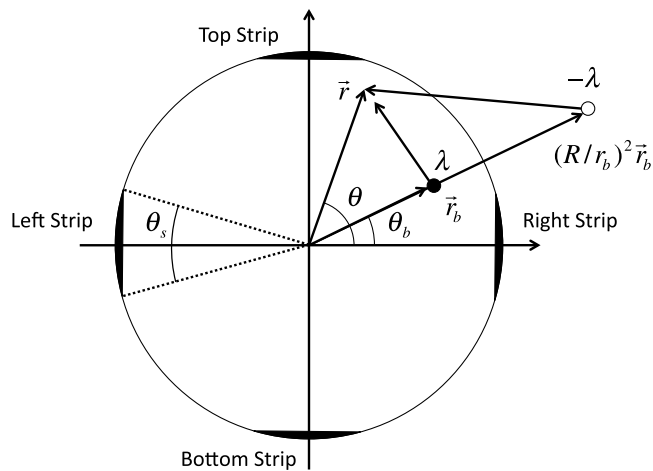

(a) Description of nonlinear BPM model parameters.

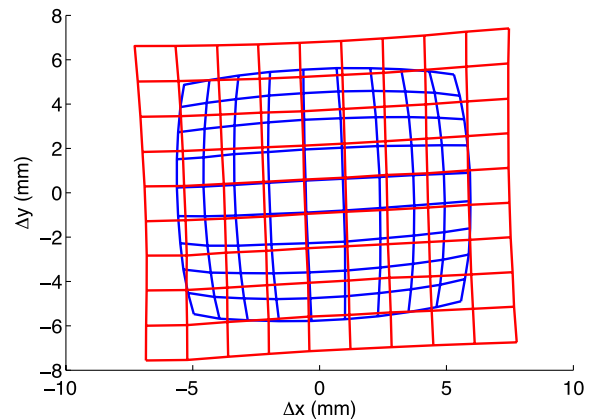

(b) Comparison of linear and nonlinear BPM models.

FIG. 6. Nonlinear BPM model description and verification: (a) shows the relevant parameters for the model, while (b) shows the comparison of the standard linear BPM position calculation (blue) and positions computed with a nonlinear correction (red) from a square grid scan of an upstream horizontal and vertical corrector pair.

In order to invert the BPM signals and obtain the beam position, the signals from Eq. (7) are fit to the injector BPM signals using a $\chi^2$ minimization with the beam position as the fit parameters. To verify this procedure, a pair of upstream horizontal and vertical corrector magnets was scanned in a grid pattern and the response on a test BPM was measured. Figure 6(b) shows the comparison of the standard linear BPM model (blue), and nonlinear model given by Eq. (7) (red). The tilt of the position grids shown in the figure is due to the rotation of the corrector magnet relative to the BPM. The inclusion of this model effectively extended the workable range of the BPMs in the injector by roughly a factor of 2. This increased range made the use of the BPMs in response measurements significantly more robust.

### 2. Difference orbits and coupler effects

To verify each injector beam line element and its corresponding GPT model, linear optics response measurements have been performed. The transverse dynamics were verified by changing the initial position of the beam on the cathode or kicking the beam with a corrector magnet and recording the change in position on all downstream BPMs. This was repeated for each type of element in the injector, starting with the gun and moving downstream turning on elements one by one and comparing the resulting response function to GPT simulations. Figure 7(a) shows an example response measurement and corresponding GPT comparison. For this measurement, the first pair of horizontal and vertical correctors in the A1 section were scanned and the response through the straight portion of the injector recorded (with all quadrupoles off). Time of flight difference orbits were also measured by adjusting the laser phase $\pm 60$ degrees relative to the cavity phases, and measuring the bunch arrival phase from all BPMs via I/Q detection and bunch signal processing. Excellent agreement with the GPT model was obtained using all BPMs, including those in the merger.

Difference orbits were also used to verify the 3D rf field maps used to model the cavities and fields near the input power couplers. Simulations show that asymmetric focusing from the couplers is worse when a low energy beam passes through the coupler fields before being accelerated [40]. Thus, to more clearly measure the effects of the couplers, we turned off all of the SRF cavities except the second one, which has couplers at the entrance of the cavity, as seen by the beam. A square grid of angles was scanned using the last pair of horizontal and vertical correctors just before the entrance to the cryomodule, and the resulting response pattern was measured on a downstream BPM. This was repeated at multiple cavity phases shifted relative to the on-crest phase. By taking the ratio of the change in position in $x$ to the change in $y$, the asymmetry in the response through the cavity was computed. Figure 7(b) shows the comparison of the $x$ to $y$ response aspect ratio measured in the injector and computed in GPT. The agreement is quite good except for the point where the response in both planes goes through zero. With these measurements, we are confident in our ability to include the 3D focusing effects of the cavity input couplers.

### 3. Alignment

Previous work has shown [8,22,25] that good alignment through each optical element is required to diminish emittance growth, and indeed alignment of the beam through the gun, emittance compensation section, and SRF cavities proved very important for obtaining the low emittance results presented here. In order to arrive at these results,

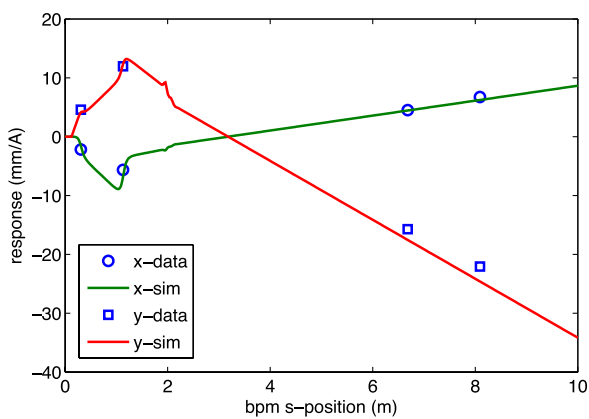
(a) Example response measurement.

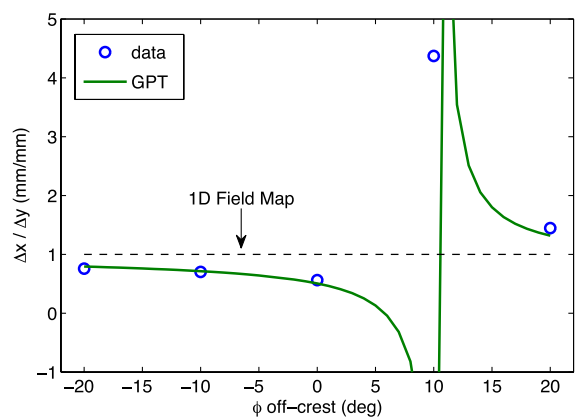
(b) Asymmetric response through the second cavity due to coupler fields.

FIG. 7. Response measurements: (a) the response from the first set of correctors through the first SRF cavity. The cavity was set to 1 MV on-crest. (b) The response asymmetry due to the coupler fields in the second cavity as a function of cavity phase. The cavity voltage was 1.5 MV. The dashed line shows the expected response from a cylindrically symmetric, or 1D field map model of the cavity.

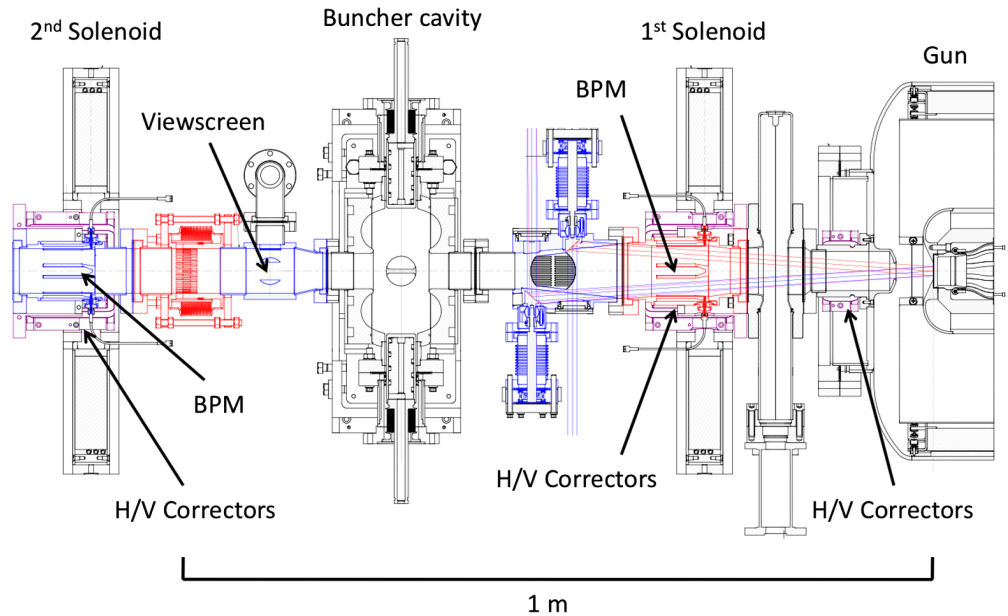

FIG. 8. Layout of the A1 emittance compensation section.

a methodical element by element alignment procedure was developed. The benefit of such an approach was that after a thorough execution of the following procedure, additional alignment work was kept to a minimum on subsequent experiments. The first step of this procedure was to center the laser spot on the cathode. To do so, the spot was scanned both horizontally and vertically to form a grid of positions. With the first solenoid off, the corresponding beam position was recorded on the viewscreen after the gun, see Fig. 8. Fitting the response data to an off-axis expansion of the gun focusing allowed us to determine the electrostatic center of the gun/cathode to within 50 $\mu$m. It should be noted that in order to achieve a good cathode lifetime, off-center laser spot operation is required to minimize ion back-bombardment [10,11]. However, we found that offsetting the laser spot by 3 mm and using a corrector pair to bring the beam back through the center of the 1st solenoid did not degrade the beam emittance by more than 5%.

After aligning to the gun, the beam was then aligned in the buncher cavity. To do so, the gun was set to 350 kV, and the first and second solenoids were degaussed and turned off. The buncher cavity was turned on at 50 kV and the two energy zero-crossing phases determined. In order to keep the transverse beam size small, the cavity phase was set to the debunching zero-crossing value, in order to provide focusing from the buncher. The use of the zero-crossing phase also eliminated the effect of dispersion due to the combination of unwanted stray fields and low beam energy. The beam position on the second viewscreen was recorded with the cavity turned off and then turned on. The initial position offset going into the cavity field region was then found by fitting the beam transfer matrix from the corrector coil pair just before the buncher to the viewscreen after the cavity. The transfer matrix was computed from the on-axis electric field map shown in Fig. 3(e) using the method derived in [40]. The position offset in the buncher was then compensated by adjusting the corrector coils just before it. Using this technique, we were routinely able to align the beam through the center of the buncher to within 20 $\mu$m.

Next, the orbit was aligned through the first two SRF cavities. Each cavity was separately turned on to 50 kV and set to the debunching zero-crossing phase just as with the buncher. Once the correct phases were found, the beam position on the A3 viewscreen was recorded for three different settings: both cavities off, and then each cavity on separately. After recording the beam position on the A3 viewscreen for each setting, the response functions from the last two pairs of horizontal and vertical correctors before the cryomodule were measured. From this set of response measurements, the corrector settings were determined that would place the beam at the same spot on the A3 viewscreen for all three cavity settings. This process produced an orbit which did not change position on the A3 viewscreen to within roughly 50 $\mu$m when the first two cavities were toggled on and off.

Finally, the solenoids were aligned. The alignment of the buncher and first two SRF cavities fixed the settings of all the available corrector coils in the A1 section. Consequently, the solenoids had to be physically moved to align their magnet centers with beam orbit. For the solenoids, both their offset and angle in the horizontal and vertical planes were found by performing a current scan of each magnet, recording the response on a downstream viewscreen, and fitting the data using the transfer matrix of the solenoid [24,41]. The physical adjustment of the solenoid positions and angles was greatly aided by the incorporation of alignment motors in the design of the solenoid magnet support structure. At the completion of the final alignment measurements, the transverse offsets of the solenoids were aligned to within roughly 50 $\mu$m, and the transverse angles to within 0.2 mrad.

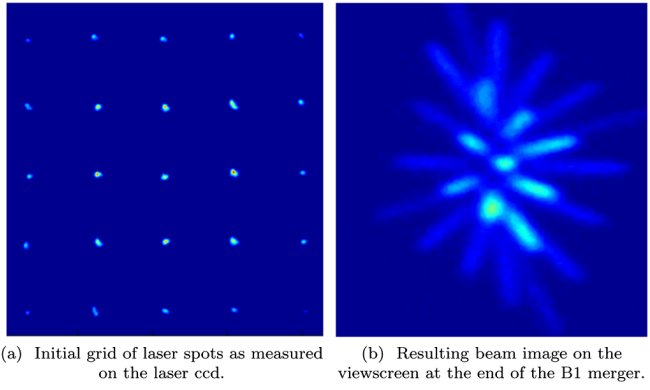

FIG. 9. Alignment check using a grid of laser spots (a) and the resulting beam image on the viewscreen at the end of the B1 merger section (b).

Alignment of the orbit through the optical elements in the A3 straight and B1 merger section was achieved by flattening the BPM readings in these sections. To check the overall alignment once the orbits for emittance measurements were set up, a special laser mask with a regular grid of 100 $\mu$m holes spaced 0.75 mm apart was placed in the laser path. Figure 9 shows the initial grid pattern and the measured grid pattern in the B1 section. To generate this image the buncher was purposefully set to give a longer

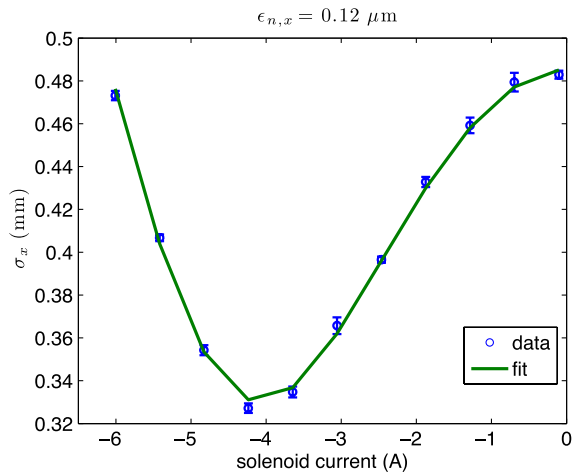

(a) Solenoid scan horizontal emittance measurements.

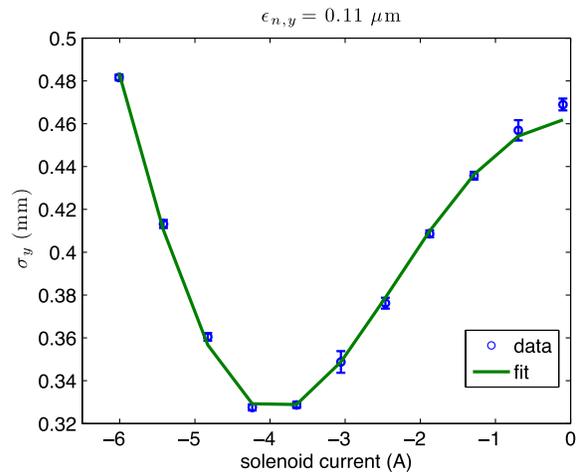

(b) Solenoid scan vertical emittance measurement.

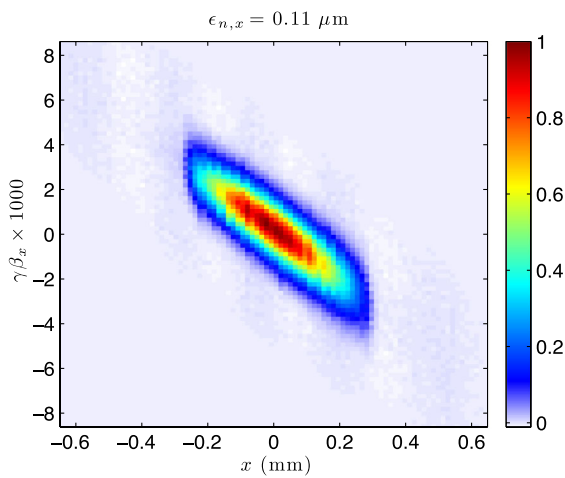

(c) Measured horizontal phase space in the merger section.

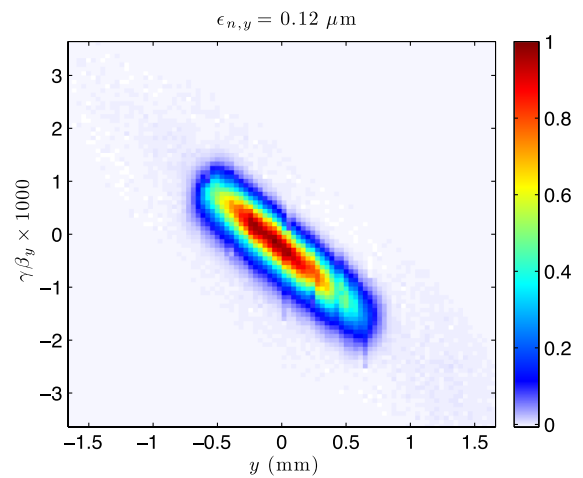

(d) Measured vertical phase space in the merger section.

FIG. 10. Projected emittance measurement at the cathode using a solenoid scan (a) and (b), and corresponding measurements in the merger section (c) and (d). The colormap and normalization in (c) and (d) is used for all subsequent phase-space plots in this work. The estimated error for these emittance values was $\pm 0.01$ $\mu$m.

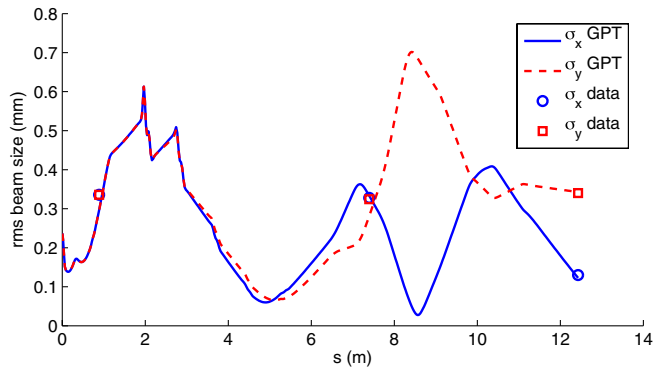

FIG. 11. Comparison of simulated and measured rms spot sizes along the beam line.

bunch length in order to exaggerate the time-dependent rf focusing from the SRF cavities. In this image, the center spot being circular and the other spots pointing towards the center indicate that the beam was aligned reasonably throughout the injector. Also, the lack of curvature to each of the spots/lines indicates that aberration effects are minimal.

### 4. Beam sizes and thermal emittance

Before measuring emittance with nonzero bunch charge, we calibrated our emittance measurement system and emittance analysis scripts by measuring the emittance at near-zero bunch charge ($q \leq 0.03$ pC). A baseline thermal emittance was measured after the gun and before the cryomodule by scanning the current of the first solenoid and measuring the beam spot size on a viewscreen downstream. By computing the linear transfer matrix through the combined gun and solenoid fields, the emittance and initial rms beam spot size were found using the method in [24]. Figures 10(a) and 10(b) show the solenoid scan data and fitted curve for the cathode used in this work. The resulting horizontal and vertical emittances measured with the solenoid current scan were $0.12 \pm 0.01$ and $0.11 \pm 0.01$ $\mu$m, respectively. To check the calibration of the EMS in the merger section, the projected horizontal and vertical phase space, as well as the horizontal time-resolved phase space were measured. For these measurements, the 19 pC/bunch injector optics settings were used (see Table II), however the bunch charge was reduced so that space charge effects were negligible. Figures 10(c) and 10(d) show the projected emittance measured in the B1 merger section with a beam momentum of roughly 8 MeV/c. The estimated systematic error in the calibration of the merger EMS system was less than 7%. The horizontal and vertical emittances from these measurements were $0.11 \pm 0.01$ and $0.12 \pm 0.01$, which agree with the solenoid scan results to within the estimated error in both measurements. The same value for the horizontal projected emittance, $0.11 \pm 0.01$ $\mu$m, was measured in the merger section using the time-resolved EMS. These measurements not only verified the EMS diagnostics and analysis procedures, but also provided an additional check of the orbit alignment.

As a final check of the optics settings in the machine and simulations, we measured the transverse rms beam sizes at several locations along the injector with near-zero bunch charge. Figure 11 shows the comparison of the simulated and measured rms spot sizes. The optics settings were the same as those used in the EMS calibration measurements. The measured values were computed from images of the beam on the A1 and A3 viewscreens, and from the phase spaces measured in the merger section shown in Figs. 10(c) and 10(d). The systematic uncertainty in these measurements due to the viewscreen calibration and setup resolution was estimated to be less than 5% for the direct viewscreen measurements. As Fig. 11 shows, excellent agreement between GPT and the measured beam sizes was found.

TABLE II. Relevant injector settings used for measurements with nonzero space charge.

| Element | Parameter | 19 pC/bunch values | 77 pC/bunch values |
| --- | --- | --- | --- |
| Laser | Pinhole (mm), intensity cutoff (%) | 1, 40 | 2, 35 |
| Laser | rms pulse length (ps) | 8 | 8 |
| DC gun | Voltage (kV) | 350 | 350 |
| Solenoid 1 | Peak field (T) | 0.032 | 0.031 |
| Buncher | Voltage (kV), phase (deg) | 50, −90 | 60, −90 |
| Solenoid 2 | Peak field (T) | −0.020 | −0.020 |
| SRF cavity 1 | Voltage (kV), phase (deg) | 1491, −10 | 1491, −10 |
| SRF cavity 2 | Voltage (kV), phase (deg) | 1953, −16 | 1953, −7 |
| SRF cavity 3 | Voltage (kV), phase (deg) | 1386, 0 | 1386, 0 |
| SRF cavity 4 | Voltage (kV), phase (deg) | 1386, 0 | 1386, 0 |
| SRF cavity 5 | Voltage (kV), phase (deg) | 1386, 0 | 1500, −20 |
| A3 quad 1 | Field gradient integral ([T/m] · m) | 0.013 | 0.013 |
| A3 quad 2 | Field gradient integral ([T/m] · m) | −0.033 | −0.033 |
| A3 quad 3 | Field gradient integral ([T/m] · m) | −0.016 | −0.016 |
| A3 quad 4 | Field gradient integral ([T/m] · m) | 0.029 | 0.029 |
| B1 quads | Field gradient integral ([T/m] · m) | −0.017 | −0.016 |

## B. Measurements with space charge

Two main data sets were produced for this work: one at 19 pC per bunch, and one at 77 pC. These correspond to 25 and 100 mA average current when operating at the full 1.3 GHz repetition rate. Each data set consists of a measurement of the projected horizontal and vertical phase spaces, the time-resolved horizontal phase space, and the energy spread distribution. All data was taken at the end of the merger section except the energy spread data, which was measured using the A4 straight section and C2 bend section. From the projected phase space, the horizontal and vertical emittance as a function of beam fraction was computed. Similarly, from the time-resolved phase space data, the slice emittance was computed as a function of beam fraction, as well as the current profile along the bunch. Refer to the Appendix for the emittance definitions used to characterize non-Gaussian phase spaces.

### 1. Injector settings and simulation parameters

To arrive at the final optics used for these experiments, optimizations of the GPT model were carried out using a multiobjective genetic algorithm [7,8]. In general, each optimization was run with two competing objectives (e.g. minimizing the emittance at the location of the merger EMS and maximizing the bunch charge), while varying the optics settings (e.g. solenoid, rf, and quad settings). Upon convergence of the optimizer, this produced an optimal front for the two objective variables. A complete list of the parameters varied in the optimizer can be found in the first and second columns of Table II. Note that for all

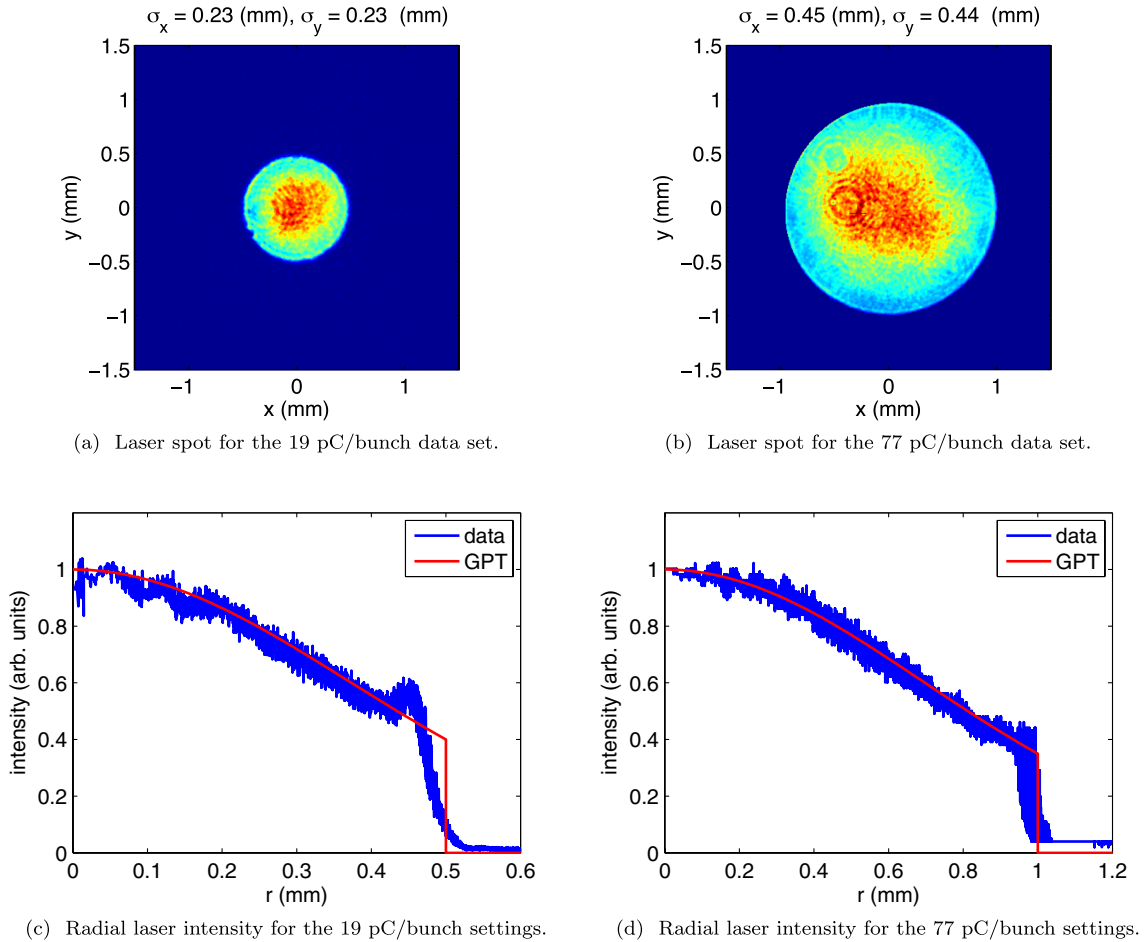

(a) Laser spot for the 19 pC/bunch data set.

(b) Laser spot for the 77 pC/bunch data set.

(c) Radial laser intensity for the 19 pC/bunch settings.

(d) Radial laser intensity for the 77 pC/bunch settings.

FIG. 12. Verification of the initial transverse laser spot in the injector and GPT input file: (a) and (b) show the measured laser spot on a CCD camera as they would appear on the cathode. The uncertainty in the rms spot sizes is (a) $\pm 0.01$ mm and (b) $\pm 0.02$ mm. The plots in (c) and (d) show the corresponding measured radial laser intensity (blue) and matching truncated Gaussian (red) used in simulation.

optimizations, the gun voltage was fixed at 350 kV, and the beam energy was constrained to be $\leq 8$ MeV to reduce neutron production from the tungsten slits in the EMS. The simulated temporal laser distribution was fixed to be roughly a flattop with 8 ps rms length, and was generated by adding 16 Gaussian pulses in accordance with the temporal laser shaping system used for the injector [26]. The transverse laser profile was a Gaussian truncated at 50% intensity with the resulting rms size varied in the optimizations.

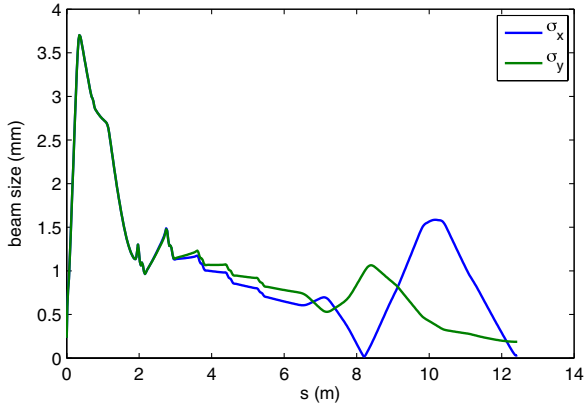
(a) RMS beam sizes for 19 pC/bunch.

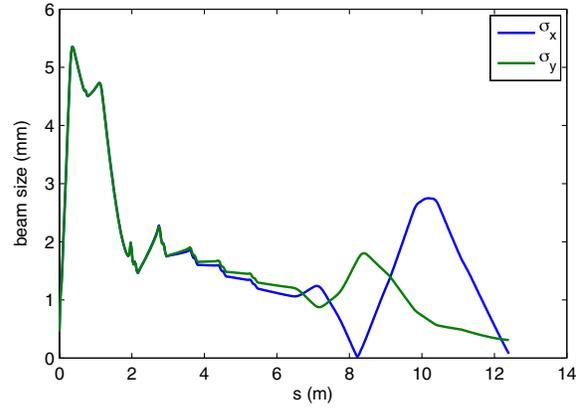
(b) RMS beam sizes for 77 pC/bunch.

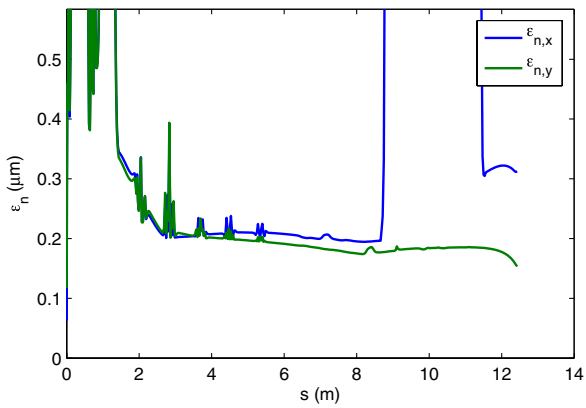
(c) Projected emittance for 19 pC/bunch.

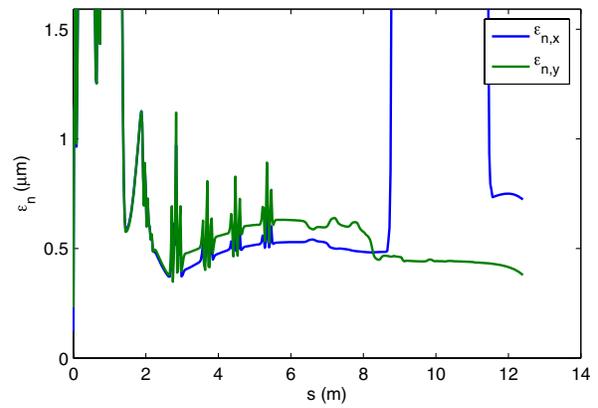
(d) Projected emittance for 77 pC/bunch.

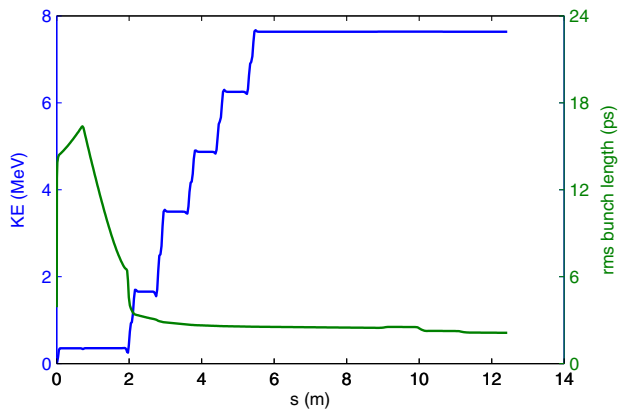
(e) Kinetic energy and bunch length for 19 pC/bunch.

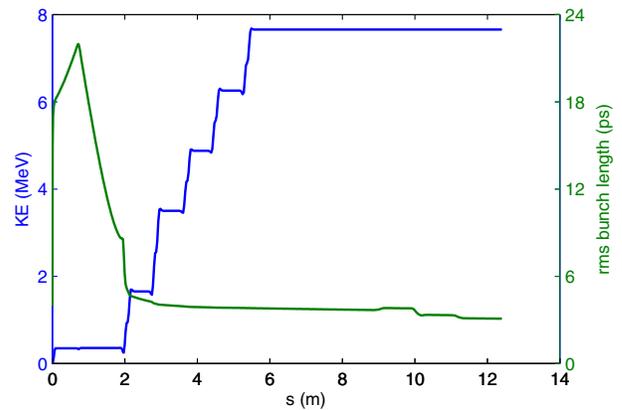
(f) Kinetic energy and bunch length for 77 pC/bunch.

FIG. 13. Simulation data for the 19 pC/bunch (left) and 77 pC/bunch (right) injector settings: (a) and (b) show the rms beam size along the injector, (c) and (d) show the projected horizontal and vertical emittances, and (e) and (f) show both the kinetic energy (left axis) and rms bunch length (right axis).

Each simulation was run with a set of constraints which ensured the physicality of the results and pushed the optimizer to explore regions of the variable space relevant to achieving the injector design goals. The two most important constraints were the rms bunch length: $\sigma_t \sim 2$ ps, and the rms energy spread $\sigma_\delta \sim 0.1$ to 0.2%. Note that these values are slightly more stringent than the quoted design goals in Table I.

Optics solutions from the last set of optimizations were loaded into the GPT virtual accelerator GUI, and then tested in the injector. Doing so led to the recognition of a common feature to all of the optimized solutions: the optimizer always focused the beam through a waist exactly at the position of the EMS in the merger. As the experiment proceeded, another general trend was observed using the virtual accelerator GUI. Optimized solutions which kept the beam sizes small, particularly in the straight section and merger, gave better measured emittance results. The settings used in the final measurements presented here were both derived from one optimization solution for 50 pC bunch charge, as this optics setting kept the beam sizes reasonably small through the entire injector. Using this parameter set as a starting point in our virtual accelerator interface, the bunch charge was reduced from 50 to 19 pC while adjusting the magnet and buncher settings to compensate for the reduced space charge effects, as well as scaling the laser spot diameter in accordance with $\sigma_{x,y} \propto \sqrt{q}$. This allowed us to keep the simulated spot sizes small through the injector, while also maintaining the location of the beam focus at the position of the merger EMS. The procedure was then started over, raising the charge from 19 to 77 pC. This time the phases of the first two SRF cavities, and the voltage and phase of the last SRF cavity were also adjusted in order to maintain small emittance values at the merger EMS.

These settings were then loaded into the injector and the measured projected emittance was minimized by scanning both solenoid currents and adjusting the intensity cutoff value in the measured transverse laser profile. Figure 12 shows the measured laser profiles used in the final measurements and the corresponding profiles used in the final GPT simulations. The final solenoid currents used in the injector were within 3% of the simulation values. The quads in the B1 section were also adjusted slightly for both optics settings, but kept within 4% of the simulations. These slight adjustments to the simulated injector settings are believed to be a consequence of hysteresis effects in the magnets, as well as error in the calibration factors used to convert machine parameters to simulation parameters. Figure 13 shows the rms beam sizes, projected horizontal and vertical emittances, kinetic energy, and bunch lengths computed using the final simulation optics values. Table II shows injector settings and parameters used in measurements. The beam kinetic energy measured after the cryomodule was 7.5 and 7.7 MeV for the two bunch charges, respectively.

### 2. Projected emittance results

As previously discussed, measuring low projected emittance after the merger section that scales according Eq. (1) and meets the design specification of the injector was one of the main goals of this work. Parts (a) and (b) of Table III show the best projected emittance data from

TABLE III. Measured and simulated projected horizontal (a), and vertical (b) emittances.

(a) Horizontal projected emittance data.

| 19 pC Measurement Type | $\epsilon_{n,x}(100\%)$ | $\epsilon_{n,x}(90\%)$ | $\epsilon_{n,x}(\text{core})$ | $f_{\text{core}}$ | $\epsilon_{n,x}(\text{core})/f_{\text{core}}$ |
|---|---|---|---|---|---|
| Projected EMS | $0.33 \pm 0.02$ $\mu$m | $0.23 \pm 0.02$ $\mu$m | $0.14 \pm 0.01$ $\mu$m | 67% | $0.21 \pm 0.01$ $\mu$m |
| Time-resolved EMS | $0.28 \pm 0.02$ $\mu$m | $0.21 \pm 0.01$ $\mu$m | $0.14 \pm 0.01$ $\mu$m | 72% | $0.19 \pm 0.01$ $\mu$m |
| GPT simulation | 0.31 $\mu$m | 0.19 $\mu$m | 0.07 $\mu$m | 59% | 0.12 $\mu$m |
| **77 pC Measurement type** | $\epsilon_{n,x}(100\%)$ | $\epsilon_{n,x}(90\%)$ | $\epsilon_{n,x}(\text{core})$ | $f_{\text{core}}$ | $\epsilon_{n,x}(\text{core})/f_{\text{core}}$ |
| Projected EMS | $0.69 \pm 0.05$ $\mu$m | $0.51 \pm 0.04$ $\mu$m | $0.28 \pm 0.2$ $\mu$m | 64% | $0.44 \pm 0.03$ $\mu$m |
| Time-resolved EMS | $0.66 \pm 0.05$ $\mu$m | $0.48 \pm 0.04$ $\mu$m | $0.29 \pm 0.2$ $\mu$m | 67% | $0.43 \pm 0.03$ $\mu$m |
| GPT simulation | 0.72 $\mu$m | 0.44 $\mu$m | 0.17 $\mu$m | 51% | 0.33 $\mu$m |

(b) Vertical projected emittance data.

| 19 pC Measurement type | $\epsilon_{n,y}(100\%)$ | $\epsilon_{n,y}(90\%)$ | $\epsilon_{n,y}(\text{core})$ | $f_{\text{core}}$ | $\epsilon_{n,y}(\text{core})/f_{\text{core}}$ |
|---|---|---|---|---|---|
| Projected EMS | $0.20 \pm 0.01$ $\mu$m | $0.14 \pm 0.01$ $\mu$m | $0.09 \pm 0.01$ $\mu$m | 70% | $0.13 \pm 0.01$ $\mu$m |
| GPT simulation | 0.16 $\mu$m | 0.11 $\mu$m | 0.06 $\mu$m | 64% | 0.09 $\mu$m |
| **77 pC Measurement type** | $\epsilon_{n,y}(100\%)$ | $\epsilon_{n,y}(90\%)$ | $\epsilon_{n,y}(\text{core})$ | $f_{\text{core}}$ | $\epsilon_{n,y}(\text{core})/f_{\text{core}}$ |
| Projected EMS | $0.40 \pm 0.03$ $\mu$m | $0.29 \pm 0.02$ $\mu$m | $0.19 \pm 0.01$ $\mu$m | 70% | $0.27 \pm 0.01$ $\mu$m |
| GPT simulation | 0.37 $\mu$m | 0.25 $\mu$m | 0.11 $\mu$m | 59% | 0.19 $\mu$m |

measurement as well as the corresponding GPT simulation values. The measured emittance data was processed with removal of a near constant background via an automatic bias determination routine similar to the methods described in [22,42]. The processed data was then used to determine the 100% beam emittance, as well as to generate the emittance vs fraction curve, defined in Eqs. (A2) and (A3), and the corresponding core emittance and core fraction, defined in Eq. (A3). These curves are shown for the horizontal and vertical projected phases at 19 (77) pC/bunch in Fig. 14. All of these procedures were automated and available to operators in the control room after each emittance measurement scan (lasting typically several seconds). Further details of the data processing and experimental procedures can be found in [43]. The measured 19 (77) pC/bunch horizontal and vertical projected 100% emittances agreed with the GPT model to within 6 (5)% and 25 (8)%, respectively. Similarly, the measured horizontal and vertical 90% emittances agreed with GPT to within 21 (16)% and 27 (16)%, respectively. We point out that the measured horizontal and vertical 100%, 90%, and core emittances obey the expected scaling law $\epsilon_n \propto \sqrt{q}$. Also of note is the fact that the horizontal core emittance for 77 pC meets the injector design specification for an ERL. In the vertical plane, both the 90% and core emittance meet this specification. For comparison purposes, Figs. 15(a) and 15(b) show the measured and simulated phase spaces after the merger for both the horizontal and vertical planes with near zero, 19, and 77 pC/bunch.

### 3. Time-resolved phase space and energy spread results

In order to satisfy the injector design requirements, it was important to verify that the emittance values were measured with an acceptable bunch length

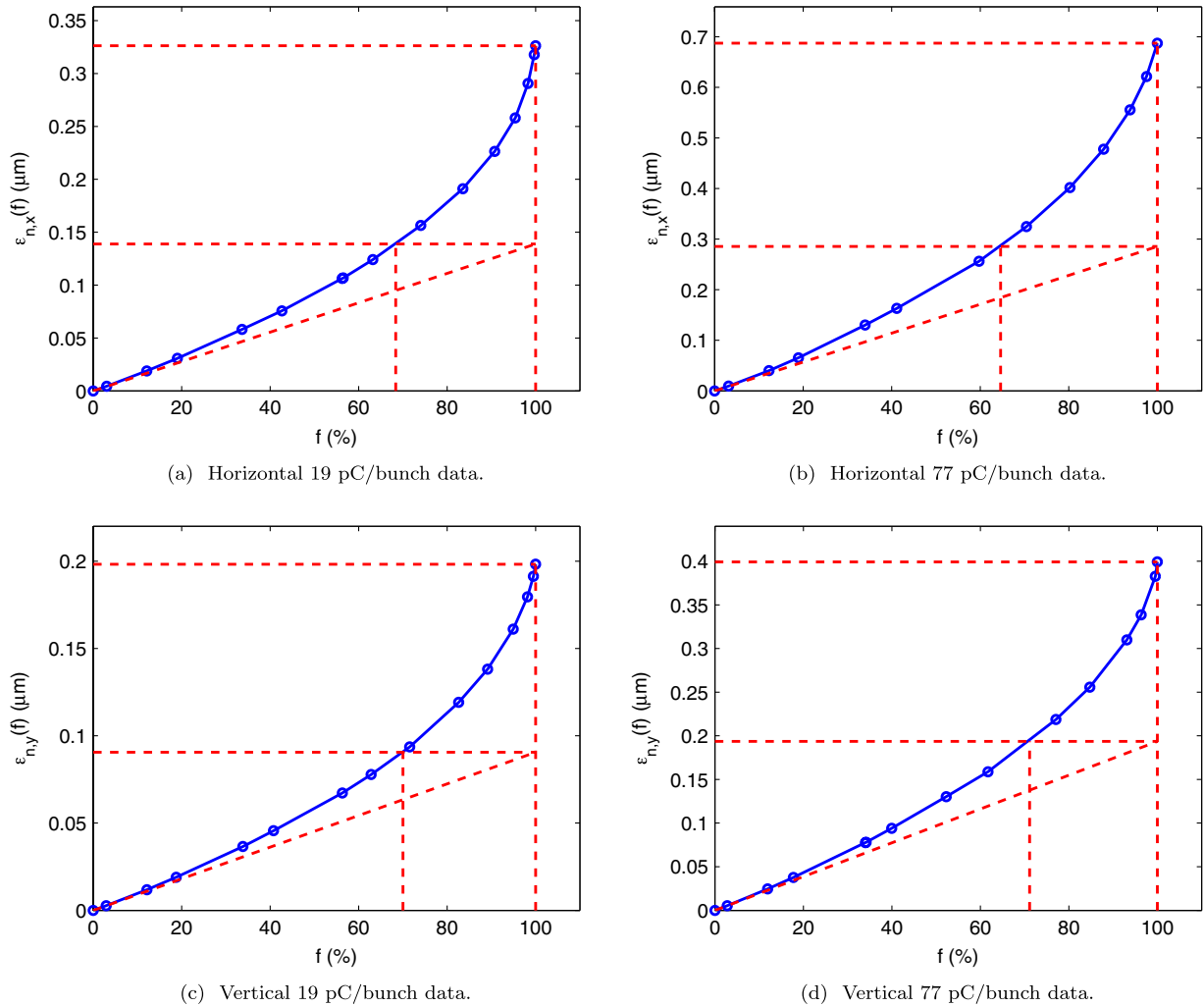

FIG. 14. Emittance versus fraction curves: (a) and (b) the curves computed from the measured horizontal phase-space data at 19 and 77 pC/bunch; (c) and (d) the curves computed from the measured vertical phase-space data at 19 and 77 pC/bunch.

($\sigma_t \leq 3$ ps). The rms bunch length was computed from the instantaneous current of each bunch measured with the time-resolved merger EMS. Figures 16(a) and 16(b) show both the measured and simulated bunch current for the 19 and 77 pC/bunch data, respectively. The rms bunch lengths for the 19 (77) pC per bunch settings were measured to be $2.1 \pm 0.1$ ($3.0 \pm 0.2$) ps, respectively, while GPT gave bunch lengths of 2.2 (3.1) ps, respectively. The agreement between measurement and GPT was within 5% in both cases. As Fig. 16(a) shows, the qualitative agreement between data and simulation was good for the 19 pC/bunch measurement. The difference in the overall scaling between the measured and simulated data for this setting is due to the normalization of the data to the bunch charge. In the 77 pC/bunch case, the qualitative agreement between

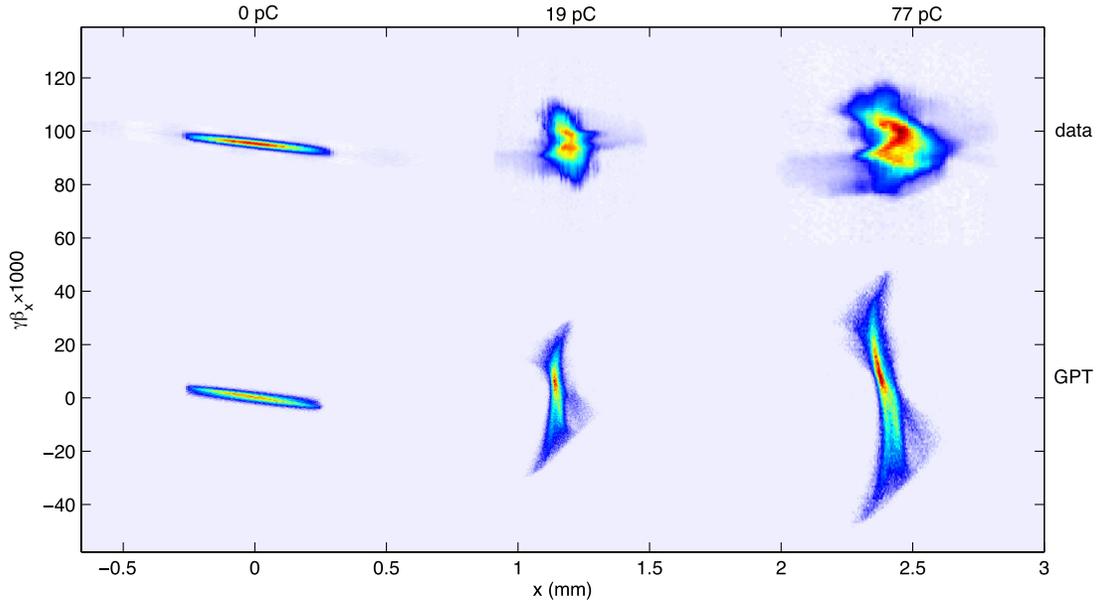

(a) Horizontal phase space as a function of bunch charge.

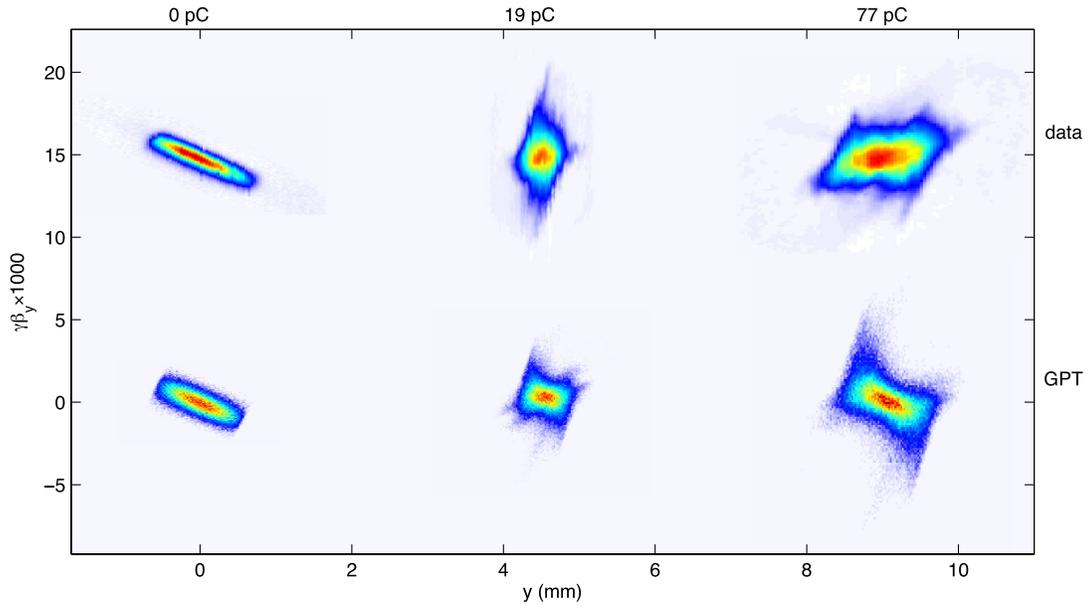

(b) Vertical phase space as a function of bunch charge.

FIG. 15. Comparison of the measured and simulated projected transverse phase space as a function of bunch charge. Plot (a) shows the horizontal phase space, while (b) shows the vertical phase space. Corresponding emittance values can be found in Table III.

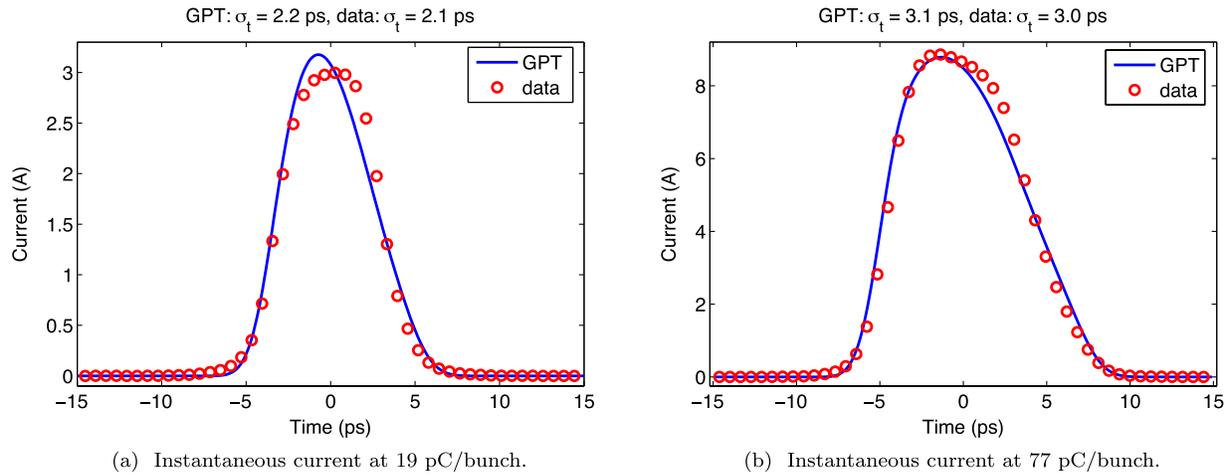

FIG. 16. Comparison of the measured beam current to GPT simulation. The estimated uncertainty in the rms bunch lengths was (a) ±0.1 ps and (b) ±0.2 ps.

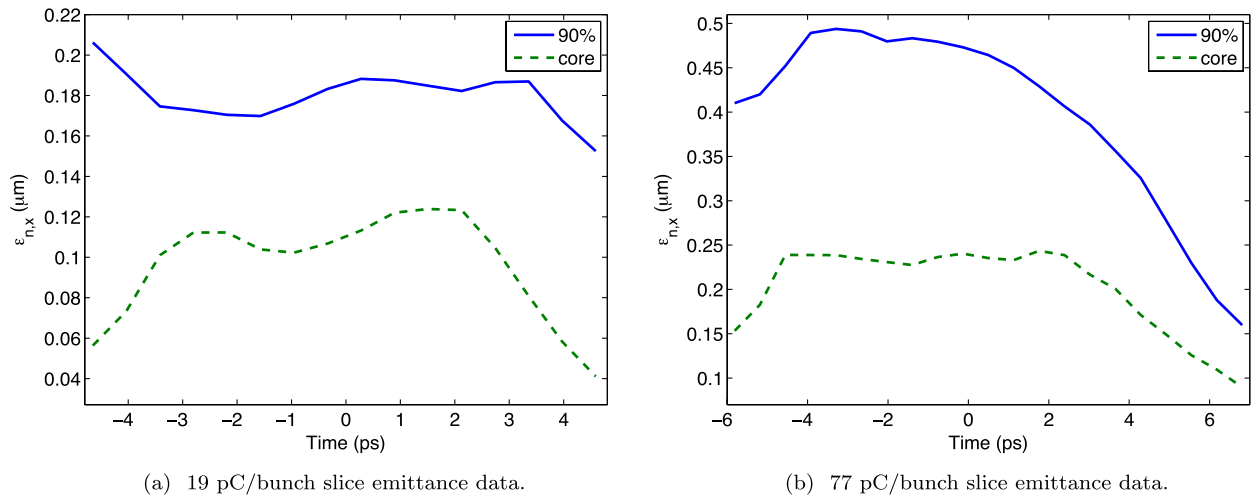

FIG. 17. The 90% and core slice emittance for (a) 19 pC/bunch and (b) 77 pC/bunch.

measurement and simulation, shown in Fig. 16(b), was excellent.

In addition to measuring the bunch length, the time-resolved emittance was measured both because it is of interest to FEL applications and to elucidate the character of the emittance growth in the merger. Figure 17 shows the core and 90% emittance for both the 19 and 77 pC measurements. For the 19 pC data, both emittances are relatively constant over the bunch length. Similarly, for the 77 pC data, the core emittance is constant over the majority of the bunch length. Also important is the fact that the core emittance for this data is below the design specification for the injector. The time-resolved emittance measurements provide a very elegant way of viewing the resulting phase-space distributions. Figure 18 shows the three-dimensional representation of the time-resolved phase space for both data sets. The 3D representation demonstrates that the z-shaped features seen in the projected emittance in Fig. 15(a) are actually a real effect formed along the time axis.

The last quantity measured was the rms energy spread. To do so, the beam was sent through the A4 straight section, followed by a single dipole and viewscreen in the C2 section (see Fig. 1). Before entering the dipole, the beam was clipped by passing it through a crossed pair of emittance measurement slits. The emittance measurement scanner magnets in this section were set so that crossed slits selected out a beamlet from the centroid of

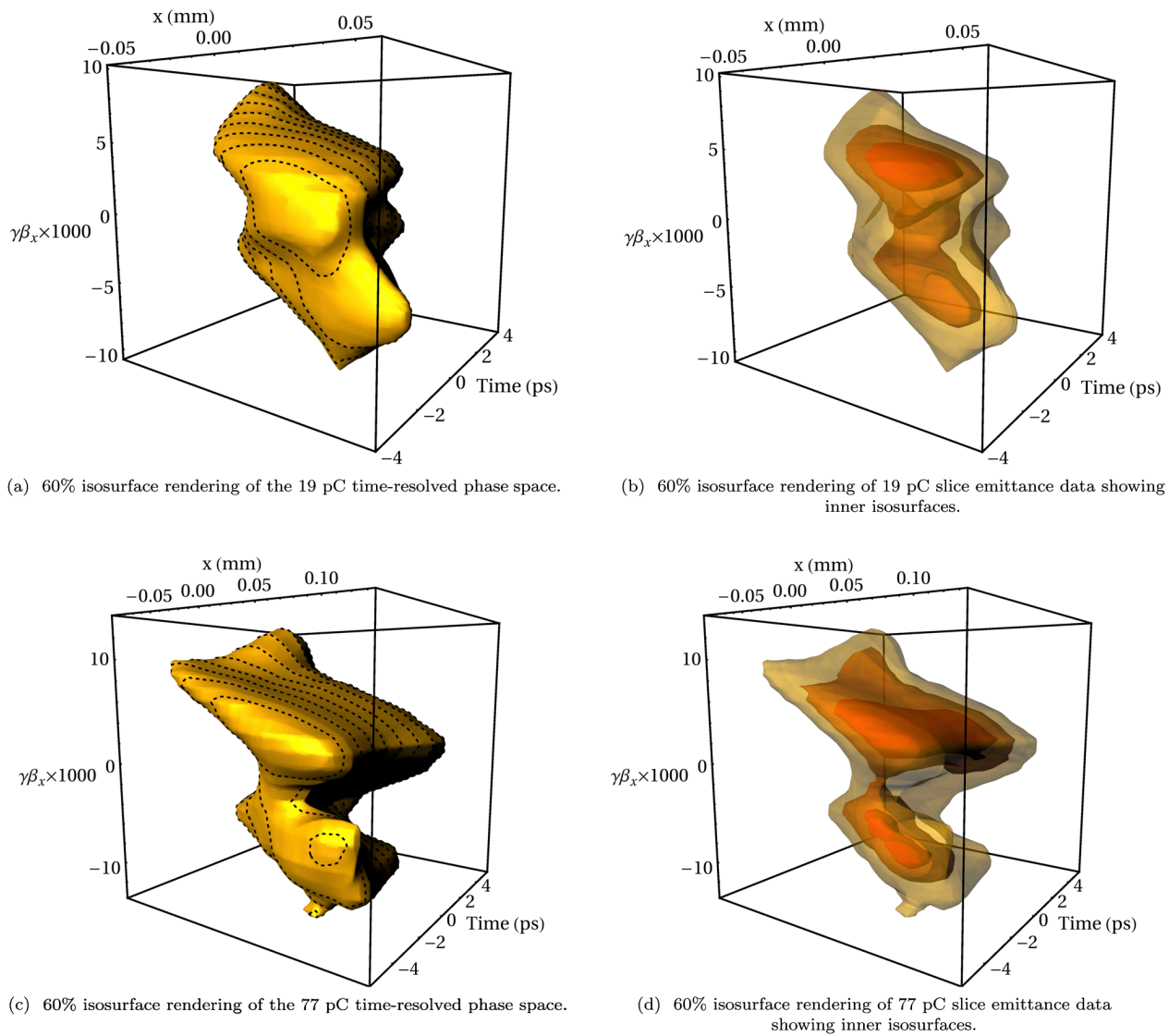

(a) 60% isosurface rendering of the 19 pC time-resolved phase space.

(b) 60% isosurface rendering of 19 pC slice emittance data showing inner isosurfaces.

(c) 60% isosurface rendering of the 77 pC time-resolved phase space.

(d) 60% isosurface rendering of 77 pC slice emittance data showing inner isosurfaces.

FIG. 18. Three-dimensional plotting of the time-resolved phase space.

the horizontal and vertical phase spaces. Table IV shows the simulated and measured rms energy spread in the straight section, as well as simulated values in the B1 merger. The measured values were computed from the 1D energy spread distribution obtained on the viewscreen in the C2 section using a 10% threshold to remove background noise. The values are slightly smaller than simulation, which is likely due to the fact that we were only measuring the energy spread of a single transverse beamlet, while the simulated values are computed from the entire beam distribution. While we did not measure the energy spread directly in the merger section, the agreement found between measurement and simulation for emittance and bunch length lead us to conclude that the values measured in the straight section at least provide an upper bound on the energy spread in the merger, following the same trend found in the simulation data.

## IV. CONCLUSION AND DISCUSSIONS

A comprehensive model of the Cornell ERL injector has been constructed using the space charge code GPT. After verifying the accuracy of the GPT model against linear optics measurements in the injector, multiobjective optimizations were carried out in order to find optics settings with which to measure low emittance after the merger section in the injector. In addition, a user interface between the GPT code, the optimizer solutions, and the injector was developed. This interface provided visualization of relevant simulation data in one-to-one correspondence with measured data, and allowed users to explore adjustments of the injector optics in simulation while in the control room, often in near real time with measurements. Using this interface, and starting from a single optimized setting of the injector model, optics sets for both 19 and 77 pC bunch charges were found which kept both the simulated rms beam sizes small throughout the injector, in addition to preserving the minimized emittance at the end of the merger section. These settings were then loaded into the injector, and the phase-space data for each bunch charge was taken.

TABLE IV. Simulated and measured rms energy spread as a function of bunch charge.

| Data type | 19 pC/bunch, A4 section | 77 pC/bunch, A4 section | 19 pC/bunch, B1 merger | 77 pC/bunch, B1 merger |
|---|---|---|---|---|
| GPT simulation | 0.16% | 0.27% | 0.12% | 0.21% |
| Measured | 0.14 ± 0.01% | 0.26 ± 0.01% | NA | NA |

The resulting data sets include the vertical and horizontal projected phase spaces, as well as the time-resolved horizontal phase space at the merger EMS, and the energy spread distribution in the A4 straight section. Overall, we found excellent agreement between measurement and simulation. For both bunch charges, the agreement between the measured projected 100 and 90% emittance values was within 30% of the simulated values in both transverse planes. We point out that for 77 pC/bunch, the measured 90% emittance in vertical plane, as well as the core emittance in both planes, meets the ERL design specification of $\epsilon_n \leq 0.3$ μm. The projected emittance in both transverse planes demonstrates the correct scaling with bunch charge shown in Eq. (1). Using the time-resolved horizontal phase data, the longitudinal bunch profile and time-resolved emittance were computed. The measured rms bunch length for both bunch charges was at or below the 3 ps specification, and agreed with simulation to within 5%. For both bunch charges, the time-resolved core emittance met the ERL specification. Finally, an estimation of the energy spread of the beam in merger was found by measuring the energy spread in the straight section. Agreement between the measured and simulated rms energy spread was within 13% for both bunch charges.

These results represent a significant advancement in high-brightness photoinjectors. The measured emittances in this work set a new record low for DC photoinjectors producing beams with comparable bunch charge. To put these results in a broader picture, it is instructive to compare the performance of the Cornell injector for its designed application of a 5 GeV x-ray ERL to the beam quality of existing storage rings. For this comparison, we assume a 100 mA, 1 nm-rad horizontal emittance storage ring with $10^{-3}$ energy spread and 1% coupling factor, representing the best of existing third generation light sources [44]. As a figure of merit for non-Gaussian beams, it is convenient to use the effective transverse average beam brightness over the rms energy spread of the beam at the location of an undulator:

$$\left( I \cdot \frac{f_x \cdot f_y}{\epsilon_x(f_x) \cdot \epsilon_y(f_y)} \bigg|_{\text{core}} \right) \times \frac{1}{\sigma_\delta}. \quad (8)$$

Here $\epsilon_x$ and $\epsilon_y$ are the transverse geometric emittance values as a function of the horizontal and vertical beam fractions, respectively. The energy spread is included in this expression to reflect the fact that undulators with larger number of periods can be more efficiently utilized for beams with smaller energy spread. In an ERL, the rms energy spread after the main linac will be defined by the rf curvature and the bunch length according to $(2\pi f_{\text{rf}} \cdot \sigma_t)^2/\sqrt{2}$ [45]. Using our 19 pC/bunch data, and assuming the full repetition rate, the estimated energy spread and effective average brightness of a 1.3 GHz, 5 GeV ERL yields a higher transverse brightness over the best storage ring by a factor of 20.

Looking forward, we point out that the measurements shown here demonstrate two crucial points: (i) that low emittances reported previously in simulations [7,8] are well within the reach of the next planned iteration of the photoinjector; and (ii) the relevant physics and control parameters required to produce these low emittances are now understood. In developing a plan for reducing the emittance further, we note that optimization results indicate that lower emittances and shorter bunch lengths at the end of the merger are possible at higher beam energies [31]. As a result, the optimal photoinjector for a future ERL light source will operate at higher beam energies (roughly 12 MeV) than those used in this work [2]. Equation (1) shows two more directions for further improvement. For a given bunch charge, the emittance in this equation can be reduced by lowering the photocathode MTE, or by increasing the accelerating field at the cathode. In fact, the results for the vertical emittance demonstrate that the emittance in this plane is dominated by the thermal emittance, and thus colder cathodes are required. Currently, there is an active cathode research program at Cornell University dedicated to improving cathode performance [46]. Already, cathodes with MTE values as low as 30 meV have been experimentally realized both for negative affinity and multialkali photocathodes [47]. In parallel, Cornell is developing an improved DC gun, in order to overcome the current voltage limitation. The new gun design features a segmented insulator with guard rings [48] in order to minimize damaging the insulator from field emission. Lastly, improved laser shaping will aid in creating bunches with more linear space charge fields. According to the rough scaling law in Eq. (1), as well as more detailed calculations reported in [2,8], these improvements are expected to reduce the emittance in the photoinjector by roughly a factor of 3, resulting in a beam brightness roughly 10 times higher than reported here. This ability to independently improve critical elements in the injector, resulting in better performance, is one of the major strengths of a linac based accelerator.

## ACKNOWLEDGMENTS

We acknowledge Mike Billing for his help with non-linear BPM corrections, as well as Bill Schaff for providing the GaAs photocathode used in these measurements. In addition, we thank Georg Hoffstaetter for his help editing this manuscript. This work is supported by National Science Foundation (Grant No. DMR-0807731), as well as the Department of Energy (Grant No. DE-SC0003965).

## APPENDIX: EMITTANCE DEFINITIONS

Here we provide the relevant emittance definitions used in this work to describe non-Gaussian phase-space distributions. We use the standard definition of the normalized transverse rms emittance:

$$\epsilon_n = \frac{1}{mc}\sqrt{\langle x^2\rangle\langle p_x^2\rangle - \langle xp_x\rangle^2}$$
$$= \sqrt{\langle x^2\rangle\langle \gamma^2\beta_x^2\rangle - \langle x\cdot\gamma\beta_x\rangle^2}, \quad (A1)$$

where $\gamma$ and $\beta_x$ are the normalized energy and transverse velocity of each electron. In this and all following expressions, the subscript "$n$" is used to distinguished between the normalized emittance $\epsilon_n$ and geometric emittance $\epsilon$, which are related by $\epsilon_n = (\gamma\beta)\cdot\epsilon$. In this and all subsequent expressions, $\langle u\rangle$ denotes the average over the particle distribution in phase space: $\langle u\rangle = \iint u(x,p_x)\rho(x,p_x)dxdp_x$, where $\rho(x,p_x)$ is the normalized 2D phase-space distribution function. The rms emittance as a function of beam fraction is defined as follows [49]. For an area in phase space $\pi a$, an ellipse with Twiss parameters given by

$$T = \begin{pmatrix} \beta_n & -\alpha_n \\ -\alpha_n & \gamma_n \end{pmatrix},$$

is defined so that the phase-space region enclosed by the ellipse is given by $d(a,T) = \{\mathbf{x}: \mathbf{x}^T T^{-1}\mathbf{x} \leq a\}$, where $\mathbf{x} = (x, p_x)^T$. The Twiss parameters in $T$ are varied until the fraction of particles enclosed in the ellipse is maximized. Labeling this phase-space region $D(a)$, the beam fraction is defined as

$$f(a) = \max\left\{\iint_{d(a,T)} \rho(x,p_x)dxdp_x\right\}$$
$$= \iint_{D(a)} \rho(x,p_x)dxdp_x. \quad (A2)$$

TABLE V. The scaled emittance and fraction data for various phase-space distributions.

| Distribution type | $\hat{\epsilon}_n(90\%)$ | $\hat{\epsilon}_n$(core) | $f_{\text{core}}$ | $\hat{\epsilon}_n$(core)$/f_{\text{core}}$ |
|---|---|---|---|---|
| Uniform | 0.90 | 1 | 100% | 1 |
| Elliptical | 0.87 | 5/6 | 87% | 0.96 |
| Gaussian | 0.74 | 1/2 | 72% | 0.69 |

The corresponding fractional emittance takes the form

$$\epsilon_n(a) = \frac{1}{mc}\sqrt{\langle x^2\rangle_D\langle p_x^2\rangle_D - \langle xp_x\rangle_D^2}, \quad (A3)$$

where $\langle u\rangle_D = \frac{1}{f(a)}\iint_{D(a)} u\rho(x,p_x)dxdp_x$. The parametric curve defined by $\{f(a),\epsilon_n(a)\}$ is the emittance vs fraction curve $\epsilon_n(f)$. Also important for understanding emittances of non-Gaussian beams are the definitions of the core emittance and corresponding core fraction [16,49]:

$$\epsilon_n(\text{core}) = \frac{d\epsilon_n}{df}\bigg|_{f\to 0}, \quad f_{\text{core}}: \epsilon_n(f_{\text{core}}) = \epsilon_n(\text{core}). \quad (A4)$$

For comparison purposes, the emittance vs fraction curves for 2D uniform, elliptical, and Gaussian distributions have

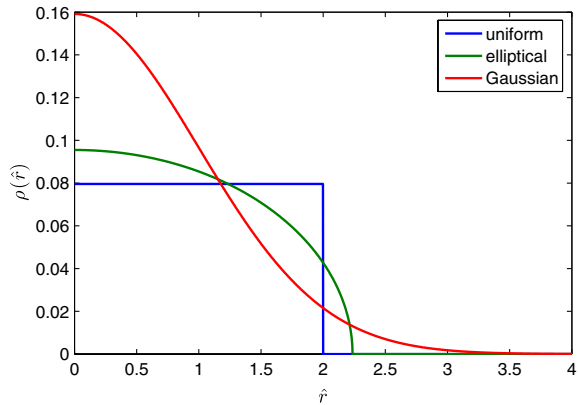
(a) Transverse phase-space distributions.

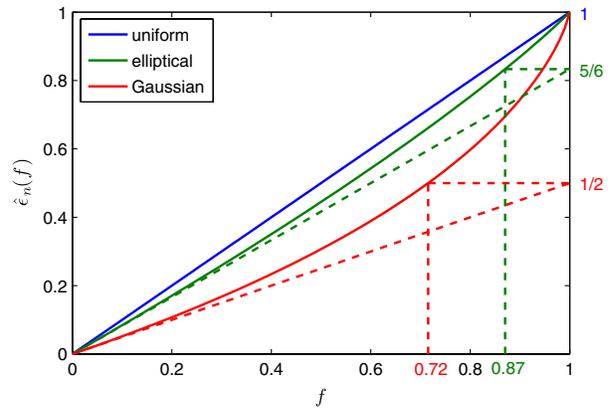
(b) Emittance verses fraction curves.

FIG. 19. Example transverse phase-space distributions as a function of the normalized coordinates $\hat{r}^2 = \hat{x}^2 + \hat{p}_x^2$ (a), and the corresponding emittance vs fraction curves (b). Dashed lines indicate core emittance and core fraction values.

been computed. To do so the correlation between $x$ and $p_x$ has been removed and the coordinates rescaled so that the distributions can be written as radial functions of the normalized coordinate $\hat{r} = \sqrt{\hat{x}^2 + \hat{p}_x^2}$. Additionally, the distributions are parametrized so that the resulting emittance vs fraction curve $\hat{\epsilon}_n(f)$ is normalized: $\hat{\epsilon}_n(f=1) = 1$. Figure 19(a) shows each of the three distributions as a function of the normalized radial coordinate. The corresponding emittance vs fraction curves are shown in Fig. 19(b). From these curves the 90% and core emittance (relative to the 100% emittance) can be computed. Table V gives these ratios, as well as the core fraction, for each distribution. For additional discussion on the connection between core emittance and brightness, see [49].

[1] G. R. Neil, C. L. Bohn, S. V. Benson, G. Biallas, D. Douglas, H. F. Dylla, R. Evans, J. Fugitt, A. Grippo, J. Gubeli, R. Hill, K. Jordan, R. Li, L. Merminga, P. Piot, J. Preble, M. Shinn, T. Siggins, R. Walker, and B. Yunn, Phys. Rev. Lett. **84,** 662 (2000).
[2] I. V. Bazarov *et al.*, Cornell University Technical Report (2012) [http://www.classe.cornell.edu/ERL/].
[3] D. H. Dowell, J. W. Lewellen, D. Nguyen, and R. Rimmer, Nucl. Instrum. Methods Phys. Res., Sect. A **557,** 61 (2006).
[4] F. Stephan *et al.*, Phys. Rev. ST Accel. Beams **13,** 020704 (2010).
[5] R. Akre, D. Dowell, P. Emma, J. Frisch, S. Gilevich, G. Hays, P. Hering, R. Iverson, C. Limborg-Deprey, H. Loos, A. Miahnahri, J. Schmerge, J. Turner, J. Welch, W. White, and J. Wu, Phys. Rev. ST Accel. Beams **11,** 030703 (2008).
[6] F. Sannibale *et al.*, Phys. Rev. ST Accel. Beams **15,** 103501 (2012).
[7] I. V. Bazarov and C. K. Sinclair, Phys. Rev. ST Accel. Beams **8,** 034202 (2005).
[8] I. V. Bazarov, A. Kim, M. N. Lakshmanan, and J. M. Maxson, Phys. Rev. ST Accel. Beams **14,** 072001 (2011).
[9] A. Arnold and J. Teichert, Phys. Rev. ST Accel. Beams **14,** 024801 (2011).
[10] B. Dunham *et al.*, Appl. Phys. Lett. **102,** 034105 (2013).
[11] L. Cultrera, S. Karkare, B. Lillard, A. Bartnik, I. Bazarov, B. Dunham, W. Schaff, and K. Smolenski (unpublished).
[12] D. H. Dowell, S. Joly, A. Loulergue, J. P. de Brion, and G. Haouat, Phys. Plasmas **4,** 3369 (1997).
[13] K.-J. Kim, Nucl. Instrum. Methods Phys. Res., Sect. A **275,** 201 (1989).
[14] B. E. Carlsten, Nucl. Instrum. Methods Phys. Res., Sect. A, **285,** 313 (1989).
[15] L. Serafini and J. B. Rosenzweig, Phys. Rev. E **55,** 7565 (1997).
[16] I. V. Bazarov, B. M. Dunham, and C. K. Sinclair, Phys. Rev. Lett. **102,** 104801 (2009).
[17] http://www.pulsar.nl/gpt/.
[18] M. Liepe, S. Belomestnykh, E. Chojnacki, Z. Conway, G. Hoffstaetter, R. Kaplan, S. Posen, P. Quigley, J. Sears, V. Shemelin, and V. Veshcherevich, in *Proceedings of the 25th International Linear Accelerator Conference LINAC10, Tsukuba, Japan* (KEK, Tsukuba, Japan, 2010), TU303, pp. 382–386.
[19] I. V. Bazarov, B. M. Dunham, Y. Li, X. Liu, D. G. Ouzounov, C. K. Sinclair, F. Hannon, and T. Miyajima, J. Appl. Phys. **103,** 054901 (2008).
[20] I. V. Bazarov, D. G. Ouzounov, B. M. Dunham, S. A. Belomestnykh, Y. Li, X. Liu, R. E. Meller, J. Sikora, C. K. Sinclair, F. W. Wise, and T. Miyajima, Phys. Rev. ST Accel. Beams **11,** 040702 (2008).
[21] I. V. Bazarov, B. M. Dunham, X. Liu, M. Virgo, A. M. Dabiran, F. Hannon, and H. Sayed, J. Appl. Phys. **105,** 083715 (2009).
[22] I. V. Bazarov, B. M. Dunham, C. Gulliford, Y. Li, X. Liu, C. K. Sinclair, K. Soong, and F. Hannon, Phys. Rev. ST Accel. Beams **11,** 100703 (2008).
[23] L. Cultrera *et al.*, Phys. Rev. ST Accel. Beams **14,** 120101 (2011).
[24] I. Bazarov, L. Cultrera, A. Bartnik, B. Dunham, S. Karkare, Y. Li, X. Liu, J. Maxson, and W. Roussel, Appl. Phys. Lett. **98,** 224101 (2011).
[25] H. Li, Ph.D. thesis, Cornell University, 2012.
[26] Z. Zhao, B. M. Dunham, I. Bazarov, and F. W. Wise, Opt. Express **20,** 4850 (2012).
[27] V. Shemelin, S. Belomestnykh, and H. Padamsee, Cornell University LNS Internal Report No. SRF 021028-08, 2008.
[28] B. Buckley and G. H. Hoffstaetter, Phys. Rev. ST Accel. Beams **10,** 111002 (2007).
[29] S. Belomestnykh, M. Liepe, H. Padamsee, V. Shemelin, and V. Veshcherevich, Cornell University ERL Internal Report No. 02-08, 2008.
[30] V. N. Litvinenko, R. Hajima, and D. Kayran, Nucl. Instrum. Methods Phys. Res., Sect. A **557,** 165 (2006).
[31] T. Miyajima and I. Bazarov, Cornell University ERL Internal Report No. 11-02, 2011.
[32] J.-G. Hwang, E.-S. Kim, and T. Miyajima, Nucl. Instrum. Methods Phys. Res., Sect. A **684,** 18 (2012).
[33] S. Belomestnykh, I. Bazarov, V. Shemelin, J. Sikora, K. Smolenski, and V. Veshcherevich, Nucl. Instrum. Methods Phys. Res., Sect. A **614,** 179 (2010).
[34] S. B. van der Geer, O. J. Luiten, M. J. de Loos, G. Pöplau, and U. van Rienen, 3D Space-Charge Model for GPT Simulations of High Brightness Electron Bunches, Institute of Physics Conf. Ser. No. 175 (2005), p. 101.
[35] G. Poplau, U. Van Rienen, B. van der Geer, and M. de Loos, IEEE Trans. Magn. **40,** 714 (2004).
[36] http://laacg1.lanl.gov/laacg/services/download_sf.phtml.
[37] http://www.cobham.com/about-cobham/aerospace-and-security/about-us/antenna-systems/kidlington/products/opera-3d.aspx.
[38] J. Wei, Y. Papaphilippou, and R. Talman, in *Proceedings of the European Particle Accelerator Conference, Vienna, 2000* (EPS, Geneva, 2000), TUP3B15.
[39] http://www.cst.com/Content/Products/MWS/Overview.aspx.
[40] C. Gulliford, I. Bazarov, S. Belomestnykh, and V. Shemelin, Phys. Rev. ST Accel. Beams **14,** 032002 (2011).
[41] C. Gulliford and I. Bazarov, Phys. Rev. ST Accel. Beams **15,** 024002 (2012).

[42] M. P. Stockli, R. F. Welton, and R. Keller, Rev. Sci. Instrum. **75,** 1646 (2004).
[43] C. Gulliford, Ph.D. thesis, Cornell University, 2013.
[44] https://photon-science.desy.de/facilities/petra_iii/machine/parameters/index_eng.html.
[45] I. V. Bazarov and G. H. Hoffstaetter, in *Proceedings of the 20th Particle Accelerator Conference, Portland, OR, 2003* (IEEE, New York, 2003), Vol. 2, pp. 842–844.
[46] L. Cultrera, I. Bazarov, J. Conway, B. Dunham, Y. Hwang, Y. Li, X. Liu, T. Moore, R. Merluzzi, K. Smolenski, S. Karkare, J. Maxson, and W. Schaff, in *Proceedings of the 3rd International Particle Accelerator Conference, New Orleans, Louisiana, USA, 2012* (IEEE, Piscataway, NJ, 2012), WEOAB02.
[47] Photocathode Physics for Photoinjectors 2012, Cornell University, Ithaca, NY, 2012 [http://www.lepp.cornell.edu/Events/Photocathode2012/WebHome.html].
[48] R. Nagai, R. Hajima, N. Nishimori, T. Muto, M. Yamamoto, Y. Honda, T. Miyajima, H. Iijima, M. Kuriki, M. Kuwahara, S. Okumi, and T. Nakanishi, Rev. Sci. Instrum. **81,** 033304 (2010).
[49] I. V. Bazarov, Phys. Rev. ST Accel. Beams **15,** 050703 (2012).